%% file: main.tex
\pdfoutput=1

\documentclass[acmsmall, screen, nonacm]{acmart}

\input{src/preamble.tex}

\title{Generative Explanations for Program Synthesizers}

\author{Amirmohammad Nazari}
\email{nazaria@usc.edu}
\affiliation{
  \institution{University of Southern California}
  \country{USA}
}

\author{Souti Chattopadhyay}
\email{schattop@usc.edu}
\affiliation{
  \institution{University of Southern California}
  \country{USA}
}

\author{Swabha Swayamdipta}
\email{swabhas@usc.edu}
\affiliation{
  \institution{University of Southern California}
  \country{USA}
}

\author{Mukund Raghothaman}
\email{raghotha@usc.edu}
\affiliation{
  \institution{University of Southern California}
  \country{USA}
}

\begin{document}

\input{src/abstract.tex}

\maketitle

\mclearpage
\input{src/intro.tex} \mclearpage
\input{src/motiv.tex} \mclearpage
\input{src/alg.tex} \mclearpage
\input{src/eval.tex} \mclearpage
\input{src/user-study.tex} \mclearpage
\input{src/limitations.tex} \mclearpage
\input{src/related.tex} \mclearpage
\input{src/concl.tex} \mclearpage

\bibliography{src/references.bib}

\clearpage
\appendix
\onecolumn
\include{src/app-prompts.tex}
\include{src/app-listings.tex}
\include{src/app-tasks.tex}

\end{document}

%% file: src/preamble.tex
\usepackage{algorithm}
\usepackage{amsmath, amsfonts, amsthm}
\usepackage{booktabs}
\usepackage{comment}
\usepackage[shortcuts]{extdash}
\usepackage[T1]{fontenc}
\usepackage{framed}
\usepackage{listings}
\usepackage{makecell}
\usepackage{longtable}
\usepackage{multirow}
\usepackage{microtype}
\usepackage{MnSymbol} 
\usepackage{paralist, enumitem}
\usepackage{subcaption}
\usepackage{tcolorbox}

\theoremstyle{plain}
\newtheorem{thm}{Theorem}[section]
\newtheorem{lem}[thm]{Lemma}

\theoremstyle{remark}
\newtheorem{note}[thm]{Note}

\usepackage{etoolbox}
\expandafter\patchcmd\csname \string\lstinline\endcsname{%
  \leavevmode
  \bgroup
}{%
  \leavevmode
  \ifmmode\hbox\fi
  \bgroup
}{}{%
  \typeout{Patching of \string\lstinline\space failed!}%
}

\newcommand{\mathlst}[1]{\lstinline|#1|}

\newcommand{\Tool}{\textsc{NomNom}}
\newcommand{\AlgP}{\operatorname{SS}}
\newcommand{\AlgPF}{\AlgP + \operatorname{F}}
\newcommand{\AlgPFR}{(\AlgPF)\lcirclearrowright}

\newcommand{\mclearpage}{}

%% file: src/abstract.tex
\begin{abstract}
Despite great advances in program synthesis techniques, they remain algorithmic black boxes.
Although they guarantee that when synthesis is successful, the implementation satisfies the
specification, they provide no additional information regarding how the implementation works or the
manner in which the specification is realized.
%
One possibility to answer these questions is to use large language models (LLMs) to construct
human-readable explanations. Unfortunately, experiments reveal that LLMs frequently produce
nonsensical or misleading explanations when applied to the unidiomatic code produced by program
synthesizers.

In this paper, we develop an approach to reliably augment the implementation with explanatory names.
We recover fine-grained input-output data from the synthesis algorithm to enhance the prompt
supplied to the LLM, and use a combination of a program verifier and a second language model to
validate the proposed explanations before presenting them to the user. Together, these techniques
massively improve the accuracy of the proposed names, from 24\% to 79\% respectively. Through a pair
of small user studies, we find that users significantly prefer the explanations produced by our
technique (76\% of responses indicating the appropriateness of the presenting names) to the baseline
(with only 2\% of responses approving of the suggestions), and that the proposed names measurably
help users in understanding the synthesized implementation.
\end{abstract}


%% file: src/intro.tex
\section{Introduction}
\label{sec:intro}


The last fifteen years have seen an explosion of work in program synthesis~%
\cite{GulwaniPolozovRishabh:SynthesisSurvey, SearchBasedProgramSynthesis}. Program synthesizers have
been applied in diverse fields, including in producing assembly code~\cite{Stoke, Chlorophyll}, in
the synthesis of network policies~\cite{Pavol:CAV2017:NetworkSynthesis}, in data wrangling and
end-user programming~\cite{FlashFill, Rishabh:BlinkFill, FlashExtract}, and in program repair~%
\cite{Qlose, Semfix, Rishabh:AutoTutor}, among others. The algorithms underlying these synthesizers
build on various ideas, including enumerative~\cite{Transit, EUSolver} and constraint-driven
search~\cite{Dillig:ConflictDrivenLearning}, version space algebras~\cite{VSA}, type-guided
enumeration~\cite{PeterMichael:PLDI2015, Nadia:POPL2019}, and prioritized search~\cite{Euphony}, and
have been shown to synthesize fairly non-trivial pieces of code.


However, there has been comparatively less attention given to the problem of helping users to
\emph{understand} this synthesized code. Although there is research on requesting additional
guidance from the user during the synthesis process, this has primarily been oriented either towards
accelerating the synthesis process itself~\cite{2020R1}, or in reducing the number of examples
needed to identify the target program~\cite{2020R4}. Notably, these approaches do not directly help
the user in understanding how the program works, and consequently, whether it accurately realizes
their intent.
\cite{SS1} have recently described a technique to characterize the desired behavior of individual
subexpressions in a synthesized program: although these \emph{subspecifications} have been shown to
help users in understanding code, they primarily take the form of logical formulas, rather than
textual comments that can be understood by less experienced programmers.


Although most synthesizers guarantee that the emitted code conforms to the specification, inductive
synthesis (i.e., from input-output examples) works from fundamentally incomplete specifications, so
there are typically multiple programs which are consistent with the provided examples~\cite{2020R2,
Roopsha:PLDI2023}. Worse still, experience has shown that writing correct and representative
specifications is difficult, and is subject both to user error and noise~\cite{2020R3,
Kupferman:Vacuity}.
It is our contention that the difficulty of writing specifications, combined with the difficulty of
understanding how the implementation works, reduces users' confidence in using automatically
synthesized code in practical applications.


One approach to produce such human-readable explanations of code is to leverage recent breakthroughs
in large language models (including models from OpenAI such as GPT-4~\cite{GPT4} and LlaMA~%
\cite{LlaMA}) and LLMs specialized for processing code (including Codex~\cite{Codex} and Copilot~%
\cite{Copilot}).

In this paper, we describe such a system. We begin with a specification\-/implementation pair
produced by an appropriate program synthesizer. We focus on implementations produced by DreamCoder~%
\cite{DreamCoder}, a recent state-of-the-art program synthesis tool, and consider the problem of
augmenting the implementation with explanatory names for its subroutines.
%
%
This problem of automatically generating meaningful names for program elements, including for
functions and identifiers has been extensively studied in software engineering research: prominent
recent examples include Code2Vec~\cite{code2vec}, JSNice~\cite{JSNice}, and
\cite{Allamanis:FSE2015}. Better chosen names have been shown to improve the readability,
maintainability, and debugging of code~\cite{ASENameSurvey}.


The main challenge with these approaches is that implementations produced by program synthesizers
are unidiomatic, leading to poor quality names being suggested by the language model / backend tool.
While researchers have investigated ideas such as scratchpads~\cite{Scratchpads} and
chain-of-thought reasoning~\cite{ChainOfThoughtPrompting} to improve the accuracy of language
models, it is not immediately clear how one can use these techniques to sequentially justify the
appropriateness of names for program elements.


The first part of our system is a technique to augment the prompt supplied to the LLM with
additional information about the specification-implementation pair. Inspired by the framework of~%
\cite{SS1}, we analyze the implementation, and recover subspecifications that describe the local
behavior of individual subroutines, thereby helping the language model to suggest more appropriate
function names.


Our second technical insight is that when the name suggested by the LLM is appropriate, then it
provides a good proxy to reason about the broader behavior of the implementation. For example,
consider the following piece of Python code which returns the last element of a non-empty list
\lstinline|l|:
\begin{lstlisting}
def last(l):
  return l[$g$(l) - 1]
def $g$(l):
  return reduce(lambda x, y: x + 1, l, 0)
\end{lstlisting}
Observe that the function $g$ actually computes the length of the list provided as input, so it is
appropriate to name the function ``\lstinline|length|''. This also suggests that any other function
which also computes the length of the list---regardless of its internal mechanics---can equivalently
play the role of $g$ in the original implementation. For example, the following function $g'$ can be
safely substituted for $g$:
\begin{lstlisting}
def $g'$(l):
  return 0 if l == [] else 1 + $g'$(l[1:])
\end{lstlisting}
This observation motivates us to use a second LLM that resynthesizes code from the proposed function
name and type signature. We substitute this resynthesized subroutine into the surrounding
implementation, and check whether the alternative implementation satisfies the global spec. The
success of this check is an effective indicator that the name is appropriate for the subroutine in
question.


In our experiments, across a set of 144~implementations produced by DreamCoder and when compared to
names written by a human expert, these two techniques raise the accuracy of the baseline LLM from
24\% to 60\% and 82\% respectively.

On the other hand, an important drawback of self-censoring mechanisms such as this is that they
drastically reduce the response rate of the system---i.e., the fraction of prompts to which it
produces a response---from 97\% to 42\%.
We mitigate this by exploiting the stochastic nature of the responses returned by the language model
and provide it with multiple opportunities to arrive at internally consistent name suggestions.
Together with other optimizations, our experiments indicate that the overall system, which we call
\Tool, produces a response to 76\% of the presented prompts, with an overall accuracy of 79\%.


Finally, we conduct two studies (with 18 student programmers each) to identify user preferences
among the names chosen by different algorithms, and to determine the impact of appropriately chosen
names on user comprehension. We find broad agreement between user preferences and the accuracy
numbers from our larger scale experiment, and that our framework boosts our measure of program
comprehension from 18\% to 76\% respectively.


\paragraph{Contributions.}

In summary, we make the following contributions in this paper:
\begin{enumerate}
\item 
  We introduce an approach to improve the transparency of program synthesizers by providing
  explanatory names to the subroutines of the implementation. 
\item 
  We develop a framework to improve the reliability of names produced by large language models
  (LLMs): The framework supplements the prompt supplied to the LLM with additional information from
  the synthesizer, and validates its output using a second LLM which resynthesizes code from the
  proposed names.
\item 
  When applied to code produced by DreamCoder~\cite{DreamCoder}, a recent state-of-the-art program
  synthesizer, and when compared to names written by a human expert, we find that our framework
  significantly improves on the accuracy of baseline approaches.
\item 
  We conduct a user study with student programmers to determine the effectiveness of our approach,
  and confirm that the proposed names help users to understand the synthesized implementation, and
  that they prefer names produced by our framework to those produced by baseline techniques.
\end{enumerate}

%% file: src/motiv.tex
\section{Overview and Motivating Example}
\label{sec:motiv}

Consider a user who wants a program that sorts a list of numbers. They may describe their intent
using input-output examples such as the following:
\begin{alignat}{1}
  f(\mathlst{[9, 2, 7, 1]}) & = \mathlst{[1, 2, 7, 9]}. \label{eq:motiv:sort-spec}
\end{alignat}
They may then realize this intent using any of a number of inductive program synthesizers~%
\cite{PeterMichael:PLDI2015, Feser:PLDI2015, Polikarpova:PLDI2016, Dillig:Burst}.

In this paper, for the sake of concreteness, we focus on programs synthesized using DreamCoder~%
\cite{DreamCoder}. Synthesis using DreamCoder runs in two phases: in the first (offline) phase, the
system uses a corpus of synthesis tasks to construct a library of reusable components (i.e.,
functions) which it then uses to more rapidly discharge the provided specification in the subsequent
(online) synthesis phase.

We adapt the specification in Equation~\ref{eq:motiv:sort-spec} from Figure~1B of~\cite{DreamCoder}.
In response, it produces a lambda term which may be transliterated into the Python code of Figure~%
\ref{fig:motiv:sort}.
Notice that the program uses non-trivial language features such as higher-order functions and that
its subroutines have uninformative sequentially-generated names (such as \lstinline|g1|,
\lstinline|g2|, \ldots). It is therefore difficult to understand how the program works, or even
confirm that it always sorts the provided list of numbers.

\begin{figure}
\begin{lstlisting}
def g2(x2):
  def g21(x21):
    def g22(x22):
      return x21 < x22
    return len(list(filter(g22, x2))) == 0
  return list(filter(g21, x2))[0]

def g1(x1):
  def g11(x11):
    def g12(x12):
      def g13(x13):
        return x12 > x13
      return x11 > len(list(filter(g13, x1)))
    return g2(list(filter(g12, x1)))
  return g11

def $f$(x1):
  def f1(x11):
    return g1(x1)(x11 + 1)
  return list(map(f1, range(len(x1))))
\end{lstlisting}
\caption{Program produced by DreamCoder that sorts a list of numbers. The top-level function is $f$.
  Equation~\protect\ref{eq:motiv:sort-spec} is an excerpt of the specification supplied to the
  synthesizer. This program has been transliterated into Python from the original lambda-expression
  which may be found in the supplementary material.}
\label{fig:motiv:sort}
\end{figure}

We also remark that the top-level auxiliary functions, \lstinline|g1| and \lstinline|g2|, correspond
to reusable components discovered by DreamCoder from the training data. The synthesizer has
therefore concluded that they are useful across a range of tasks, and it appears plausible that they
perform some high-level conceptually salient operations over lists. Several recent program
synthesizers, including Babble~\cite{Babble} and Enumo~\cite{Enumo}, similarly attempt to learn
libraries of reusable components / rewrite rules.

Upon reflecting on this program, one may conclude that invoking the function \lstinline|g1(l)(n)|
produces the \lstinline|n|-th smallest element of the list \lstinline|l|, and that the function
\lstinline|g2| returns the largest element of the list \lstinline|x2| that it accepts as input. In
fact, in the original example of~\cite{DreamCoder}, the authors manually add expository comments
describing the behavior of these intermediate functions.
In this section, we provide an overview of our system \Tool: it accepts as input a
specification-implementation pair $(\varphi, f)$ such that $f$ satisfies $\varphi$, and uses an LLM
to algorithmically produce names for each subroutine $g$ that appears in $f$.


\mclearpage
\subsection{The Baseline LLM}
\label{sub:motiv:baseline}

As a baseline, one may request a large language model, such as one from the GPT family, to provide a
name for each function $g$ in question. Each prompt includes the body and type of the function
$g : T$ being named, and (recursively) any auxiliary functions in the call graph rooted at $g$. We
provide the baseline prompt templates in the supplementary material.

However, when using GPT-3.5, it consistently fails to produce appropriate names for any of the
functions in Figure~\ref{fig:motiv:sort}. As an example, it suggests the name
``\lstinline|largestSmallestIndices|'' for the top-level function $f$, and
``\lstinline|findNearestNumber|'' and ``\lstinline|getFirstItemMinThanArgumentValue|'' for the
functions \lstinline|g1| and \lstinline|g2| respectively.

This is somewhat unsurprising, because the program in Figure~\ref{fig:motiv:sort} is unidiomatic
Python code. In fact, if one were to replace the function bodies for \lstinline|g1| and
\lstinline|g2| with the more conventional:
\begin{lstlisting}
def g1(l):
  def g11(n):
    return sorted(l)[n]
  return g11
\end{lstlisting}
and
\begin{lstlisting}
def g2(l):
  return max(l)
\end{lstlisting}
respectively, then the system produces accurate names for each function:
\lstinline|get_sorted_values| for the top-level function $f$, and \lstinline|get_nth_sorted_element|
and \lstinline|calculateMax| for \lstinline|g1| and \lstinline|g2| respectively.

The poor performance of baseline name suggestion techniques is not limited to LLMs: as we will see
in Section~\ref{sec:eval}, even Code2Vec~\cite{code2vec}, a state-of-the-art graph embedding-based
name generation tool, produces poor quality names when applied to such unidiomatic code.

As we observe in our user study in Section~\ref{sec:user-study}, nonsensical and misleading names
massively inhibit program comprehension, and diminish the user's confidence in future predictions
from the system. In this context, the central challenges that we address in this paper are:
\begin{inparaenum}[(\itshape a\upshape)]
\item How do we provide additional information to the language model in order to guide it towards
  better-chosen names? And
\item can we validate the names produced by the system before presenting them to the user?
\end{inparaenum}
We will describe our solution to these challenges in the rest of this section.


\mclearpage
\subsection{Prompt Expansion Using Subspecifications}
\label{sub:motiv:subspecs}

Recall that our computational problem is to produce a name for each subroutine $g$ that appears in
the implementation $f$. The first part of our solution involves providing additional information to
the language model about the role of $g$ in the operation of $f$.

For example, one might conceptually extend the function \lstinline|g2| with instructions to log its
execution:
\begin{lstlisting}
def g2(x2):
  ...
  ans = list(filter(g21, x2))[0]
  print(f'g2({x2}) = {ans}') # <-- Instrumentation
  return ans
\end{lstlisting}
Note that the original specification is presented in the form of input-output examples, so one can
mechanically evaluate them to confirm that the implementation satisfies the spec. Upon testing the
implementation with the logging code enabled, one finds that:
\begin{alignat}{1}
  \mathlst{g2}(\mathlst{[1]})          & = \mathlst{1}, \label{eq:motiv:subspecs:g2:1} \\
  \mathlst{g2}(\mathlst{[2, 1]})       & = \mathlst{2}, \label{eq:motiv:subspecs:g2:21} \\
  \mathlst{g2}(\mathlst{[2, 7, 1]})    & = \mathlst{7}, \text{ and} \label{eq:motiv:subspecs:g2:271} \\
  \mathlst{g2}(\mathlst{[9, 2, 7, 1]}) & = \mathlst{9}. \label{eq:motiv:subspecs:g2:9271}
\end{alignat}
These observations immediately suggest that the function \lstinline|g2| is computing the largest
element of the list that it takes as input. While this does not provide conclusive proof, careful
reading of the code confirms this hypothesis. In addition, one may conclude that \emph{any} function
$\mathlst{g2}'$ which satisfies Equations~\ref{eq:motiv:subspecs:g2:1}--%
\ref{eq:motiv:subspecs:g2:9271}, regardless of whether or not it is otherwise semantically
equivalent to \lstinline|g2|, can be substituted into the original implementation of Figure~%
\ref{fig:motiv:sort} without affecting the fact that $f$ satisfies the global specification,
Equation~\ref{eq:motiv:sort-spec}.

This motivates us to extend the prompt supplied to the LLM with local input-output behavior of the
function $g$ being named. For example, for the function \lstinline|g2|, we use the extended prompt
shown in Figure~\ref{fig:motiv:template}.

\begin{figure}
\begin{tcolorbox}
The implementation satisfies the specification. Choose a meaningful name for the function
``\lstinline|g2(x: List[int]) -> int|'':

\noindent
Specification:
\begin{alignat*}{1}
  \mathlst{g2}(\mathlst{[1]})          & = \mathlst{1}, \\
  \mathlst{g2}(\mathlst{[2, 1]})       & = \mathlst{2},  \\
  \mathlst{g2}(\mathlst{[2, 7, 1]})    & = \mathlst{7}, \text{ and}  \\
  \mathlst{g2}(\mathlst{[9, 2, 7, 1]}) & = \mathlst{9}.
\end{alignat*}

\noindent
Implementation:
\begin{lstlisting}
def g2(x2):
  def g21(x21):
    def g22(x22):
      return x21 < x22
    return len(list(filter(g22, x2))) == 0
  return list(filter(g21, x2))[0]
\end{lstlisting}
\end{tcolorbox}
\caption{Example prompt for name generation when extended with local input-output
  subspecifications.}
\label{fig:motiv:template}
\end{figure}

With new information of this kind, the LLM is able to choose a more appropriate name for
\lstinline|g2|: \lstinline|findLargestElement|. It also manages to recover the intent of the
top-level function, $f$, for which it suggests the name \lstinline|sortList|. In our experiments in
Section~\ref{sec:eval}, when applied to the list processing benchmarks solved by DreamCoder,
providing logs of input-output behavior measurably improves the accuracy of names suggested by the
LLM from 24\% to 60\% respectively.

\begin{note}
The behavior of the function \lstinline|g2|, as described by Equations~\ref{eq:motiv:subspecs:g2:1}%
--\ref{eq:motiv:subspecs:g2:9271}, is closely related to the concept of subspecifications recently
introduced by~\cite{SS1}. The major difference is that while subspecifications are necessary and
sufficient conditions that characterize alternative implementations, monitoring input-output
behavior merely provides sufficient conditions: there might conceivably be alternative
implementations $\mathlst{g2}'$ that that violate Equations~\ref{eq:motiv:subspecs:g2:1}--%
\ref{eq:motiv:subspecs:g2:9271} but which would nevertheless result in
the global specification, Equation~\ref{eq:motiv:sort-spec} being satisfied. Regardless, we will
adopt their terminology, and refer to these logs of input-output behavior as the subspecifications
of individual subroutines.
\end{note}

The primary technical difficulty in formalizing and obtaining these subspecs is the presence of
higher-order functions. For example, naively instrumenting the function \lstinline|g1| would
produce outputs of the form:
\begin{alignat*}{1}
  \mathlst{g1}(\mathlst{[9, 2, 7, 1]}) & = \mathlst{<function g11 at 0x...>}.
\end{alignat*}
This output arises from the difficulty in serializing closures and higher-order functions. Our
solution in Section~\ref{sub:alg:subspecs} will involve a new specially designed interpreter to
recover subspecifications for higher-order functions, yielding the result:
\begin{alignat}{1}
  \mathlst{g1}(\mathlst{[9, 2, 7, 1]})(1) & = \mathlst{1}, \label{eq:motiv:subspecs:g1:1} \\
  \mathlst{g1}(\mathlst{[9, 2, 7, 1]})(2) & = \mathlst{2}, \label{eq:motiv:subspecs:g1:2} \\
  \mathlst{g1}(\mathlst{[9, 2, 7, 1]})(3) & = \mathlst{7}, \text{ and} \label{eq:motiv:subspecs:g1:3} \\
  \mathlst{g1}(\mathlst{[9, 2, 7, 1]})(4) & = \mathlst{9}, \label{eq:motiv:subspecs:g1:4} 
\end{alignat}
which immediately suggests that evaluating the function \lstinline|g1(l)(n)| produces the
\lstinline|n|-th smallest element of the list \lstinline|l|.

Unfortunately, even with this new information, in the run we consulted while writing this paper, the
language model still suggested incorrect names for \lstinline|g1| (the suggested name is
``\lstinline[literate={-}{}{0\discretionary{-}{}{-}}]|genNextGreater-Value|'') and the other
subroutines in the implementation. Note however that responses from language models are inherently
stochastic, so a subsequent run might not exactly reproduce these observations. In fact, our final
implementation in \Tool{} makes productive use of this non-determinism.


\mclearpage
\subsection{Algorithmic Sanity Checks}
\label{sub:motiv:filter}

Our next insight is that when a function is appropriately named, that name can be used to
substantially recover the original implementation. For example, recall that GPT-3.5 suggested the
name \lstinline|findLargestElement| for the function \lstinline|g2|. Given this proposed function
name and its type, we can request a second language model to reproduce the corresponding function,
to which it responds:
\begin{lstlisting}
def findLargestElement(x: List[int]):
  maxValue = 0
  for val in x:
    if maxValue < val:
      maxValue = val
  return maxValue
\end{lstlisting}
Observe that this resynthesized implementation, $\mathlst{g2}' = \mathlst{findLargestElement}$, is
not semantically equivalent to the original function \lstinline|g2|. In particular, it does not fail
on empty inputs and it also assumes that the list does not contain any negative numbers. Despite
these differences, observe that $\mathlst{g2}'$ continues to satisfy the same subspecification in
Equations~\ref{eq:motiv:subspecs:g2:1}--\ref{eq:motiv:subspecs:g2:9271}. It can therefore be
substituted into the larger implementation of Figure~\ref{fig:motiv:sort} without affecting overall
correctness, i.e., Equation~\ref{eq:motiv:sort-spec}. This suggests that
\lstinline|findLargestElement| is indeed an appropriate name for the subroutine \lstinline|g2|.

Conversely, recall that with the extended prompt of Section~\ref{sub:motiv:subspecs}, the language
model suggested the name \lstinline|GenNextGreaterValue| for \lstinline|g1|. Once again, we might
ask the LLM to produce an alternative implementation of a function with this name, and type
\lstinline|List[int]| \lstinline|->| \lstinline|Callable[[int], int]|. We provide a listing of its
output in the supplementary material. Unsurprisingly, the implementation does not satisfy the
subspecification corresponding to \lstinline|g1|, i.e., Equations~\ref{eq:motiv:subspecs:g1:1}--%
\ref{eq:motiv:subspecs:g1:4}.

Although such checks do not guarantee the appropriateness of names, they at least confirm some
degree of internal consistency. This gives us a mechanism to detect and filter out inappropriate
names. In fact, in the run we consulted while writing this paper, this \emph{algorithmic sanity
check} provides support for the proposed names \lstinline|findLargestElement| and
\lstinline|sortList| for the functions \lstinline|g2| and $f$ respectively. It also successfully
refuted the spurious name suggestions for the remaining functions in the implementation.


Overall, in our experiments in Section~\ref{sec:eval}, this further boosts the accuracy of the name
suggestions from 60\% to 82\% respectively. On the other hand, notice that this technique is
essentially a censor that filters out inappropriate names, whose use reduces the \emph{response
rate} from 97\% to 42\% respectively. One of the final optimizations in our system, \Tool, is a
technique to exploit the non-determinism in LLM outputs and regenerate names upon failure of the
sanity check. This manages to recover the drop in response rate from 42\% to 72\%, albeit with a
slight decrease in accuracy from 82\% to 79\% respectively.


\begin{note}
Given the extensive use of LLMs in our approach, the reader might wonder whether:
\begin{inparaenum}[(\itshape a\upshape)]
\item it might be possible to directly use the language model to synthesize code, and completely
  bypass the use of the underlying program synthesizer, and
\item whether the alternative implementation $\mathlst{g}'$ produced by the language model might
  somehow be more idiomatic and appropriate for presentation to the user.
\end{inparaenum}
While this is certainly possible, this approach would sacrifice guarantees inherited from the
underlying synthesizer, including that the implementation $f$ satisfies the provided input-output
examples. In addition, many synthesis tasks are formulated in the context of a target DSL, and there
is no guarantee that an implementation produced by a language model would follow the syntactic
constraints of the target DSL.
\end{note}

\begin{figure}
\includegraphics[width=.6\columnwidth]{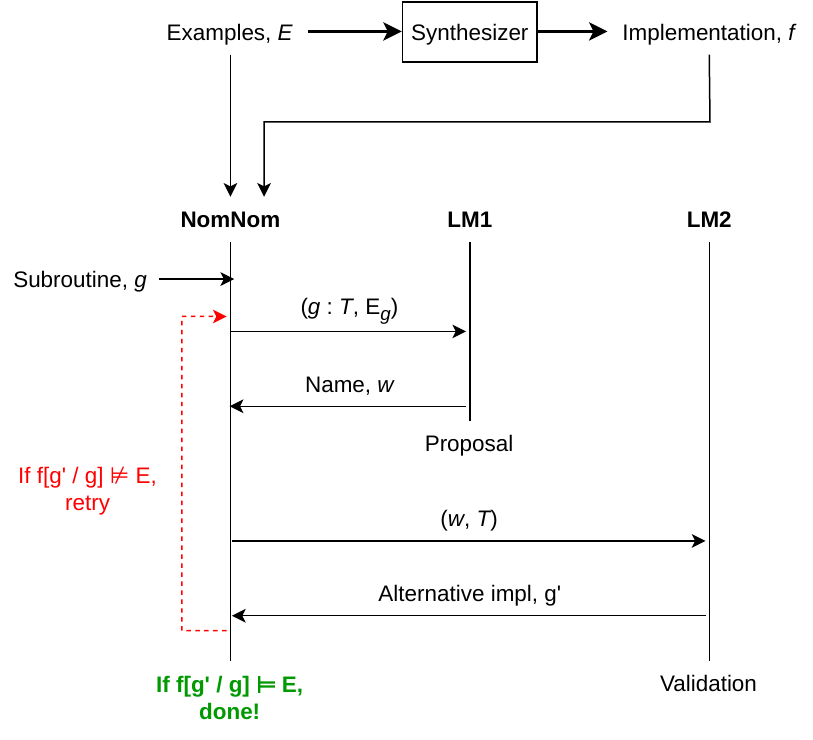}
\caption{Overall architecture of \Tool. We begin with a specification-implementation pair, $(E, f)$,
  and a subroutine $g$ of interest. The system alternates between querying a first LLM to obtain
  proposals $w$ for $g$ and validating $w$ by resynthesizing an alternative implementation $g'$
  using a second language model.}
\label{fig:motiv:arch}
\end{figure}

We describe the overall architecture of \Tool{} in Figure~\ref{fig:motiv:arch}. We devote the next
section to discussing its underlying algorithms.

%% file: src/alg.tex
\section{Algorithmic Name Synthesis}
\label{sec:alg}

We devote this section to describing our algorithm and some optimizations. The user starts by
providing a set of input-output examples, $E = \{ (i_1, o_1), (i_2, o_2), \dots, (i_n, o_n) \}$.
These examples may be drawn from integers, Boolean values, and lists of values. Upon successful
synthesis, DreamCoder returns a program $f$ which satisfies the specification $E$, i.e., for all
$(i, o) \in E$, $f(i) = o$. We indicate this by writing $f \models E$. The implementations produced
by DreamCoder are expressed as lambda-terms, examples of which may be found in the supplementary
material. Given this specification\-/implementation pair, our system proposes meaningful names for
each subroutine $g$ appearing in $f$. We present the top-level procedure in Algorithm~\ref{alg:alg}.

\begin{algorithm}
\caption{$\Tool(E, f, g)$. Given a set of input-output examples $E = \{ (i_1, o_1), (i_2, o_2),
  \dots, (i_n, o_n) \}$, implementation $f \models E$, and a subroutine $g$ of $f$, produces a name
  $w$ for $g$.}
\label{alg:alg}
\begin{enumerate}
\item Compute the local subspecification for $g$, $E_g = \{ (i_{g1}, o_{g1}), (i_{g2}, o_{g2}),
  \dots \}$.
\item Repeat until retries are exhausted:
  \begin{enumerate}
  \item \label{enu:alg:alg:p} ($\AlgP$.) Request a name $w$ for $g$ by supplying its type $g : T$
    and the subspecification $E_g$ and by using the prompt template from
    Figure~\ref{fig:user-study:template2}.
  \item \label{enu:alg:alg:q2} Request an alternative implementation $g' : T$ of a function named
    $w$ by using the prompt template from Figure~\ref{fig:user-study:template3}.
  \item \label{enu:alg:alg:pf} ($\AlgPF$.) Substitute the new implementation $g'$ into $f$. If
    $f[g' / g] \models E$, then return the name $w$.
  \end{enumerate}
\item Report failure.
\end{enumerate}
\end{algorithm}

We begin by performing a best-effort analysis of the program, and associate each sub-expression $t$
of the program with a type $t : T$. In addition, while presenting these programs to the user and to
the language model, we freely alternate between their representations as lambda-terms and as
programs expressed in a restricted subset of Python. For example, we present the lambda-term
\lstinline|(lambda (x) (cons (+ x 1) (cons (+ x 2) nil)))| to users as follows, inventing
placeholder identifiers as needed:
\begin{lstlisting}
def f(x):
  return [ x + 1, x + 2 ]
\end{lstlisting}

We now discuss the two principal elements of the algorithm, namely prompt expansion using
subspecifications and the subsequent algorithmic sanity checks.


\mclearpage
\subsection{Prompt Expansion Using Subspecifications}
\label{sub:alg:subspecs}

As discussed in Section~\ref{sub:motiv:subspecs}, obtaining the local input-output examples $E_g$
for prompt expansion is conceptually simple: one can place instrumentation code at appropriate
points inside the function body, and log the inputs and outputs being sent into and produced by the
function $g$ currently being named. The hope is that the local input-output behavior provides clues
to the overall purpose of $g$ that is not apparent from its function body. However, this procedure
is tricky because of the presence of higher-order functions. In particular, the function $g$ might
either itself take a function (closure) as input, or produce a closure as output, or possibly even
both. We develop a custom interpreter to address these challenges.

First, our interpreter includes a mechanism to print (serialize) closures. Thus, for example, the
subspecification of \lstinline|g1| in the following program:
\begin{lstlisting}
def g1(h):
  return h(3)
def $f_1$(x):
  return g1(lambda y: x + y)
\end{lstlisting}
given the global input-output example $f_1(\mathlst{2}) = 5$ is given by:
\begin{alignat*}{1}
  \mathlst{g1}(\mathlst{lambda y: 2 + y}) & = \mathlst{5}.
\end{alignat*}
Notice that this constraint is satisfied by the original subroutine \lstinline|g1|, but is also
satisfied by other implementations, including by the constant-valued function $\mathlst{g1}' =
\mathlst{lambda y: 5}$. Furthermore, it is safe to use any such new implementation $\mathlst{g1}'$
instead of the previous implementation \lstinline|g1|.

Next, we consider the case when subroutines themselves return closures. Consider, for example, the
subroutine \lstinline|g2| in the following program:
\begin{lstlisting}
def g2(x):
  def h(y): return x + y
  return h
def $f_2$(x):
  return g2(x)(3)
\end{lstlisting}
With the global input-output example $f_2(\mathlst{2}) = \mathlst{5}$, the subroutine \lstinline|g2|
is invoked with the input \lstinline|x = 2|. If we instrument the return value of \lstinline|g2| as
before, we would observe that $\mathlst{g2}(\mathlst{2}) = \mathlst{lambda y: 2 + y}$. Notice that,
given an alternative implementation $\mathlst{g2}'$, it is conceptually difficult to compare
closures for equality.

We instead use the following procedure. Say we wish to produce the subspec for a function
\lstinline|g| which itself produces a closure as output. In this case, given the original argument
$\mathlst{x}_1$, we wrap the closure $\mathlst{g}(\mathlst{x}_1)$ in a monitor object $m_1 =
([\mathlst{x}_1], \mathlst{g}(\mathlst{x}_1))$. We incorporate successive arguments to $m_1$, say
$\mathlst{x}_2$, $\mathlst{x}_3$, \ldots, $\mathlst{x}_k$, resulting in the monitor object $m_k =
([\mathlst{x}_1, \mathlst{x}_2, \dots, \mathlst{x}_k],
\mathlst{g}(\mathlst{x}_1)(\mathlst{x}_2)(\cdots)(\mathlst{x}_k))$. When the closure finally reduces
to a ground value, the monitor $m = (l, v)$ prints the sequence of function arguments $l$ and the
finally produced ground value $v$, which are then included as part of the subspecification $E_g$.

In the example above, recall that the global specification was $f_2(\mathlst{2}) = \mathlst{5}$.
Querying the subspecification for \lstinline|g2| initially results in the monitor object $m_1 =
([ 2 ], \mathlst{lambda y: 2 + y})$. Further evaluation using this monitor object produces the final
subspec $\mathlst{g2}(2)(3) = 5$. We can show that:
\begin{lem}
Let $E$ be a set of input-output examples, and $f \models E$ be a conformant implementation. Let $g$
be a subroutine in $f$, with local input-output subspecification $E_g$. Pick a function $g'$ such
that $g' \models E_g$. Then $f[g' / g] \models E$.
\end{lem}


\mclearpage
\subsection{Algorithmic Sanity Checks}
\label{sub:alg:filter}

The names proposed by the LLM in response to the query in Step~\ref{enu:alg:alg:p} are sometimes
directly embedded in the original source code, or are presented with some other decoratory text,
such as ``\emph{Name: \guillemotleft name\guillemotright}''. We have devised a set of simple
extractor routines and regular expressions that detect these patterns and appropriately extract the
proposed name. We hope to simplify this process by using structured prompting techniques in future
versions of the system~\cite{LMP}.

We then forward the proposed name $w$ and the type $T$ of the subroutine $g$ being named to a second
language model using the prompt template from Figure~\ref{fig:user-study:template3}. We interpret
the response from the LLM as an alternative implementation $g'$ of the original subroutine $g$. This
step might fail either because the response from the LLM is not a syntactically well-formed program,
or if it fails to have the desired type $T$, or if substituting it into the surrounding
implementation compromises the overall correctness specification, $f[g' / g] \not\models E$. In any
of these cases, we reject the name $w$ being proposed in response to the naming query in Step~%
\ref{enu:alg:alg:p}. Note that we only generate names for top-level subroutines, so we do not have
to consider the possibility of variable capture. This assumption greatly simplifies our
implementation.

Finally, if $f[g' / g] \models E$, then we certify the name $w$ as having passed the sanity check.


\mclearpage
\subsection{Optimizations}
\label{sub:alg:opt}

Finally, our implementation in \Tool{} includes two optimizations which increase the overall
response rate of the system without losing accuracy.


\paragraph{Retries.}

It turns out that the algorithmic sanity checks of Section~\ref{sub:alg:filter} are very effective
in discovering inconsistencies between functions and their proposed names. Filtering names using
this heuristic therefore massively improves the accuracy of the naming algorithm. Unfortunately,
this accuracy improvement is accompanied by a corresponding drop in the number of queries
successfully answered, as we will see in Section~\ref{sec:eval}.
%
The non-deterministic responses generated by language models provide an easy approach to mitigate
this drop. When a proposed name fails the sanity check, we repeatedly retry (with a limit of 20~%
attempts) until the check succeeds, leading to the outermost loop in Algorithm~\ref{alg:alg}.


\paragraph{Bottom-up name generation.}

Finally, there are certain functions which prove to be difficult to name even after multiple
independent queries. One example is the following function $f$:
\begin{lstlisting}
def $f$(x1):
  return a2(x1)(5)
def a2(x2):
  def a21(x21):
    def a22(x22):
      return x22>0
    return a22(countOccurrences(x2)(x21))
  return a21
\end{lstlisting}
Observe first that calling the function \lstinline|a2(l)(n)| tests the output of
\lstinline|countOccurrences| to determine whether the value \lstinline|n| occurs in the list
\lstinline|l|. It therefore follows that calling the top-level function $f(l)$ checks whether the
provided list contains an occurrence of the number \lstinline|5|.

The last optimization in \Tool{} facilitates this reasoning process by iteratively finding names for
higher-level functions only after all lower-level functions, i.e., those reachable from it in the
call graph have been successfully named.\footnote{Note that the concepts learned by DreamCoder
naturally have a hierarchical structure in the form of a DAG.} In our experiments, this turns out to
cause a slight increase in the response rate of the system, including for the function $f$ above.
This optimization has a flavor similar to emerging techniques for prompting language models such as
scratchpads and chain-of-thought reasoning~\cite{Scratchpads, ChainOfThoughtPrompting}.

%% file: src/eval.tex
\section{Experimental Evaluation}
\label{sec:eval}

Our implementation of \Tool{} uses \lstinline|text-davinci-003| as our backend language model. We
use the default language model settings for name generation, and only change \lstinline|max_tokens|
to 1,000 for the reverse code generation pass. Our evaluation focuses on the following research
questions:
\begin{enumerate}[label=\textbf{RQ\arabic*.}, ref=RQ\arabic*, leftmargin=*]
\item \label{enu:eval:effectiveness} How effective is our system in producing well-chosen names for
  subroutines?
\item \label{enu:eval:rr} How frequently does the system produce suggestions for subroutine names?
\item \label{enu:eval:numq} How many queries does the system require in order to propose these
  names?
\end{enumerate}


\paragraph{Benchmarks.}
Our evaluation dataset started with 155~specifications involving list processing programs which were
synthesized by DreamCoder.\footnote{File named
\lstinline|jobs/list_hard_test_ellisk_2019-02-15T11.43.28| from the DreamCoder artifact~%
\cite{DreamCoder:Artifact}.} Each of these specifications was associated with a name, indicating the
user's intent, and a varying number of implementations (from 1 to 16).
Upon manually inspecting these implementations, we discovered that 3 of them did not satisfy the
stated user intent, and 8 implementations which we were unable to explain. We eliminated these
programs, and chose the largest remaining implementation for each specification, which left us with
a dataset consisting of 144~specification-implementation pairs and which consisted of a total of
344~subroutines.
We manually provided reference names for each of these subroutines. The appropriateness of these
names was subsequently validated by a visiting student researcher who was not among the authors of
this paper.

\paragraph{Baselines.}
In addition to the baseline LLM and our algorithmic variants, we also evaluated the performance of
Code2Vec~\cite{code2vec}. Because Code2Vec works with Java code, we translated each of the benchmark
programs into Java by hand. We will include these implementations in our artifact.




\mclearpage
\subsection{\ref{enu:eval:effectiveness}: Effectiveness of Explanations}
\label{sub:eval:effectiveness}

\begin{figure}
\centering
\includegraphics[width=.6\columnwidth]{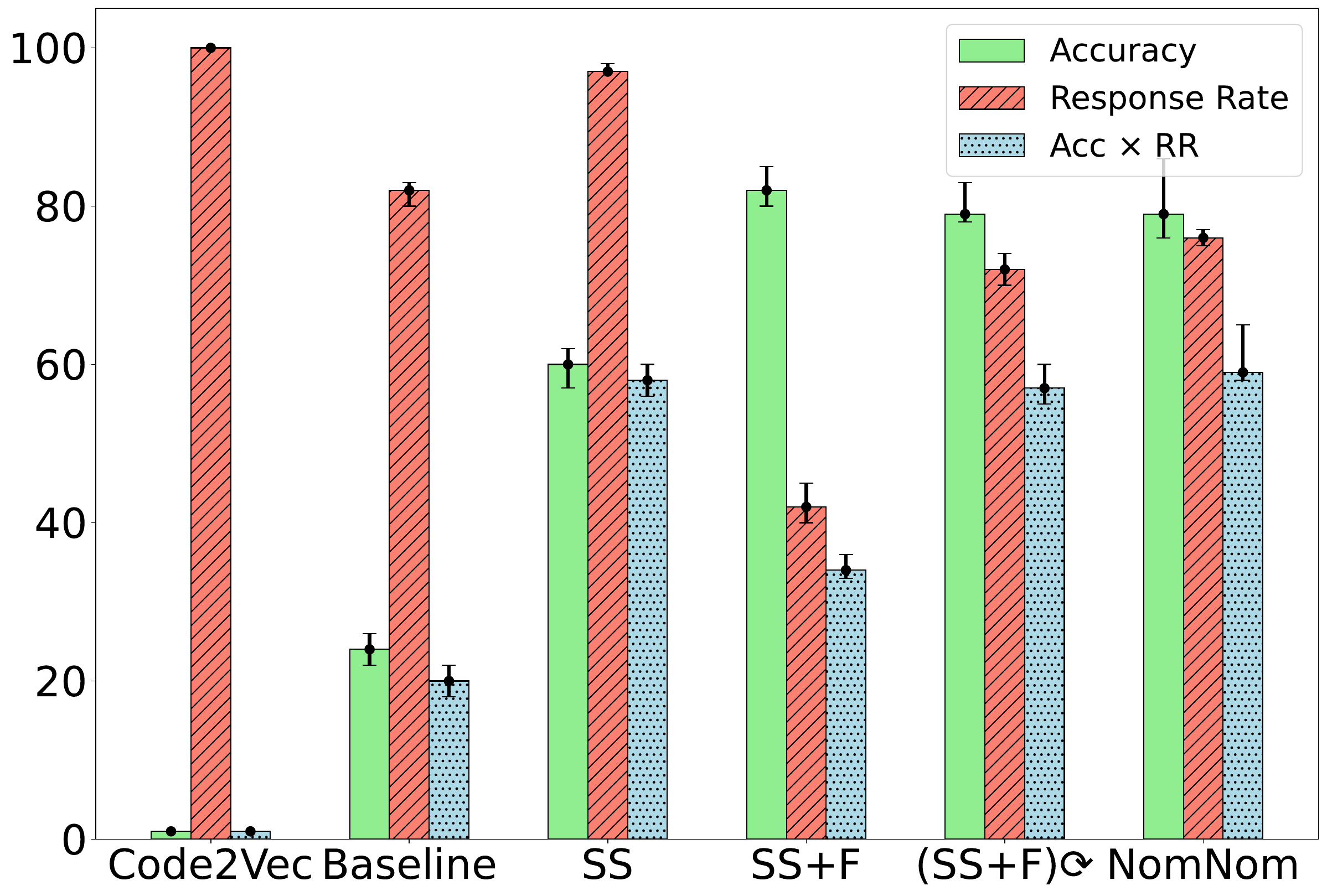}
\caption{Effectiveness of the algorithmic variants. $\AlgP$ indicates prompt expansion
  with local input-output subspecifications, $\AlgPF$ indicates the subsequent algorithmic sanity
  check, $\AlgPFR$ indicates the version which repeatedly retries upon failure, and \Tool{}
  indicates our final system with bottom-up name generation. The bars represent the median of five
  independent executions.}
\label{fig:eval}
\end{figure}

In order to measure the effectiveness of \Tool{} in producing evocative function names, we ran five
variants of our algorithm across the subroutines in our evaluation dataset. We then compared the
algorithmically produced function names to our reference names and computed the Jaro similarity
between the two~\cite{advancesInRecordLinkageMethodology, comparisonOfStringMetrics}. We declared
the proposed name to be appropriate if this similarity measure exceeded 0.7. Finally, in order to
estimate the variability of the overall procedure, we ran each algorithm five times for each
subroutine. We present our observations in Figure~\ref{fig:eval}.

Notice the consistent improvement in accuracy as we incorporate algorithmic improvements from a
baseline score of 24\% to the final value of 79\%. One example of a successfully named function is
the following:
\begin{lstlisting}
def a2(x2):
  def a21(x21):
    def a22(x22):
      return x2[x22]
    return list(map(a22, range(x21)))
  return a21
\end{lstlisting}
Evaluating the function \lstinline|a2(l)(n)| returns the first \lstinline|n| elements of the list
\lstinline|l|. Only the last two algorithms consistently suggest names such as
\lstinline|getFirstNItems|, \lstinline|selectFirstElements|, etc., while $\AlgP$ and $\AlgPF$
occasionally produce appropriate names. On the other hand, without any additional information, the
baseline LLM suggests the misleading name, \lstinline|transformList|.

Also, observe that Code2Vec generates poor quality names: this is because our benchmarks make
heavy use of higher-order functions, resulting in unidiomatic Java code. In many cases, in the
absence of an alternative, the system simply regurgitates the original placeholder function name,
\lstinline|f1|, \lstinline|f2|, etc.

One concern with our evaluation methodology might involve the validity of the reference names. Of
the 40~proposed function names polled in our user study, we only observed 3~names which we thought
were appropriate (with sufficiently high Jaro similarity to the reference), but for which the
average score of the users was ``Neutral'' or less.
Conversely, we did observe situations where we marked a proposed name as inappropriate, even though
users subsequently thought otherwise. One example is the following function \lstinline|c2|:
\begin{lstlisting}
def c2(x3):
  def c31(x31):
    def c32(x32):
      return x32==x31
    return len(list(filter(c32, x3)))
  return c31
\end{lstlisting}
Evaluating the function \lstinline|c2(l)(k)| counts the number of occurrences of \lstinline|k| in
\lstinline|l|. Our reference name was \lstinline|countOccurrencesOfK|, while $\AlgPFR$ suggested the
name \lstinline|countElementsMatchingValue|, which we rejected based on an insufficient Jaro
similarity to the reference name.

\mclearpage
\subsection{\ref{enu:eval:rr}: Response Rate of Tools}
\label{sub:eval:rr}

Although the algorithmic sanity check, $\AlgPF$, significantly improves the accuracy of the overall
algorithm, one concern is that suppressing responses from the LLM might lead to the overall system
answering a smaller number of queries. We therefore measured the response rate of the system, and we
include this data in Figure~\ref{fig:eval}.

First, observe that providing additional information to the language model, i.e., going from the
baseline algorithm to one with prompt expansion, $\AlgP$, modestly increases the overall response
rate from 82\% to 97\% respectively. The real benefit of prompt expansion comes from the massive
increase in the number of queries correctly answered, $\text{Acc} \times \text{RR}$, from 20\% to
58\%.

Next, we observe that the greater accuracy of $\AlgPF$ is accompanied by a corresponding drop in
response rate from 97\% to 42\% respectively. Indeed, it is not possible for a filtering pass to
increase the total number of queries which are correctly answered, so that the product, $\text{Acc}
\times \text{RR}$, actually experiences a drop upon its application. We also provide the confusion
matrix from one run of $\AlgPF$ in Table~\ref{tab:eval:rr}. The overall F1 score of the filter turns
out to be 0.68, so better filter designs is an important direction of future work.

\begin{table}
\caption{Confusion matrix from one run of $\AlgPF$. We compare the names produced by the LLM to our
  reference names, and declare a match when the Jaro similarity exceeds 0.7.}
\label{tab:eval:rr}
\centering
\begin{tabular}{
  >{\centering}m{0.25\columnwidth}%
  >{\centering}m{0.3\columnwidth}%
  >{\centering}m{0.3\columnwidth}}
\toprule
  \textbf{Reference} & \textbf{Filter Approves} & \textbf{Filter Rejects} \tabularnewline
\midrule
  Match     &  116 / 344  &  82 / 344  \tabularnewline
  Mismatch  &  25 / 344  &  121 / 344  \tabularnewline
\bottomrule
\end{tabular}
\end{table}

Lastly, the figure also confirms the need for the final two algorithmic variants, $\AlgPFR$ and
\Tool: By giving the system multiple opportunities to produce an internally consistent response,
they somewhat restore the response rate and provide modest increases in the product measure from
34\% to 57\% and 59\% respectively.

\mclearpage
\subsection{\ref{enu:eval:numq}: Number of LLM Queries Used}
\label{sub:eval:numq}

Finally, we measured the number of LLM queries needed by the different algorithmic variants to name
each function. We list these statistics in Table~\ref{tab:eval:numq}. Both the baseline approach and
the variant with prompt expansion, $\AlgP$, require just one LLM query to produce their response,
while the version with algorithmic sanity checks enabled, $\AlgPF$, needs two queries: the first to
produce a name suggestion and the second to reverse-synthesize the subroutine body. On the other
hand, the last two variants, with retries enabled, need to make additional queries when the first
query either fails to elicit a response or receives a response which fails validation. Note that we
report the median number of queries in Table~\ref{tab:eval:numq} as the average is skewed by
subroutines for which we hit the limit of 20~retries and eventually fail to produce a name. 

\begin{table}
\caption{Number of LLM queries needed by the algorithmic variants to name each subroutine. We report
  the median over five independent runs.}
\label{tab:eval:numq}
\centering
\begin{tabular}{
  >{\centering}m{0.4\columnwidth}%
  >{\centering}m{0.4\columnwidth}}
\toprule
  \textbf{Algorithm} &
  \textbf{Num Queries} \tabularnewline
\midrule
  Baseline &
    1 \tabularnewline
  $\AlgP$ &
    1 \tabularnewline
  $\AlgPF$ &
    2 \tabularnewline
  $\AlgPFR$ &
    4 \tabularnewline
  \Tool &
    4 \tabularnewline
\bottomrule
\end{tabular}
\end{table}

While we did not explicitly track the time needed to name each subroutine or the cumulative cost of
LLM queries, the statistics in Table~\ref{tab:eval:numq} provide some guidance. Note that the time
needed to name each subroutine is dominated by the response time from the OpenAI servers, and
depends on numerous other factors such as load on the LLM implementation. In our experience, the
most resource-intensive algorithms, $\AlgPFR$ and \Tool, produce responses within 5--10~seconds for
each subroutine.

%% file: src/user-study.tex
\section{User Study}
\label{sec:user-study}

To determine whether names help users in understanding the outputs of program synthesis
tools, we conducted two user studies to answer the following questions:
\begin{enumerate}[label=\textbf{RQ\arabic*.}, ref=RQ\arabic*, leftmargin=*, resume]
\item \label{enu:user-study:purpose} Do names help users in inferring the top-level purpose of each
  subroutine?
\item \label{enu:user-study:relation} Do names help users in understanding subroutines and
  the relationships between them?
\item \label{enu:user-study:prefs} How do user preferences vary among the names produced by
  different algorithmic variants?
\end{enumerate}


\paragraph{Participant selection.}
After IRB approval, we recruited 36 students who were familiar with Python from the engineering
schools (Computer Science, Electrical Engineering, Mechanical Engineering, and Materials
Science departments) of 7~prominent American and Canadian universities. These participants had
different levels of experience in programming and were a mix of undergraduate, Masters', and Ph.D.
students. We posit that the variation in experience is representative of users of program synthesis
tools. We randomly divided the 36 participants into two groups with 18 participants for each user
study.


\mclearpage
\subsection{Tasks and Study Structure}
\label{sub:user-study:structure}

Before each user study, we had a short screening quiz asking participants to write
a Python program that computes the sum of the elements in a list. Disregarding minor syntactic
errors, all participants passed the screening quiz. We then showed participants a short video
describing the tasks they needed to complete, and a brief introduction to aspects of the Python
language that would be heavily used, including higher-order functions and some syntactic quirks.
After the study was complete, we had a short discussion with each participant. During the
discussion, participants gave their feedback about which aspects of the study they found easy or
difficult, and their experience while answering questions. The study materials may be found in
Appendices~\ref{app:listings} and~\ref{app:user-study}.



\paragraph{Study~1: Explanations.}
We chose four programs from the larger dataset of 88 implementations produced by DreamCoder so that
there was no mutual overlap in the subroutines used. In the first study, we asked participants to
examine these programs and explain how they worked. We first asked participants to explain what each
subroutine did, and then we asked them to walk us through the execution of the program on a specific
input.
We conducted the study in one of two randomly chosen conditions, with names suggested either by the
baseline language model, or our final bottom-up name synthesizer, \Tool. We ensured that each
participant attempted two tasks with names from the baseline approach, and the remaining two tasks
with names from our system. In every case, the names were directly embedded inside the program.

\begin{figure*}
  \hfill
  \subfloat[\label{fig:user-study:explanations:time}]{
    \includegraphics[width=.3\textwidth]{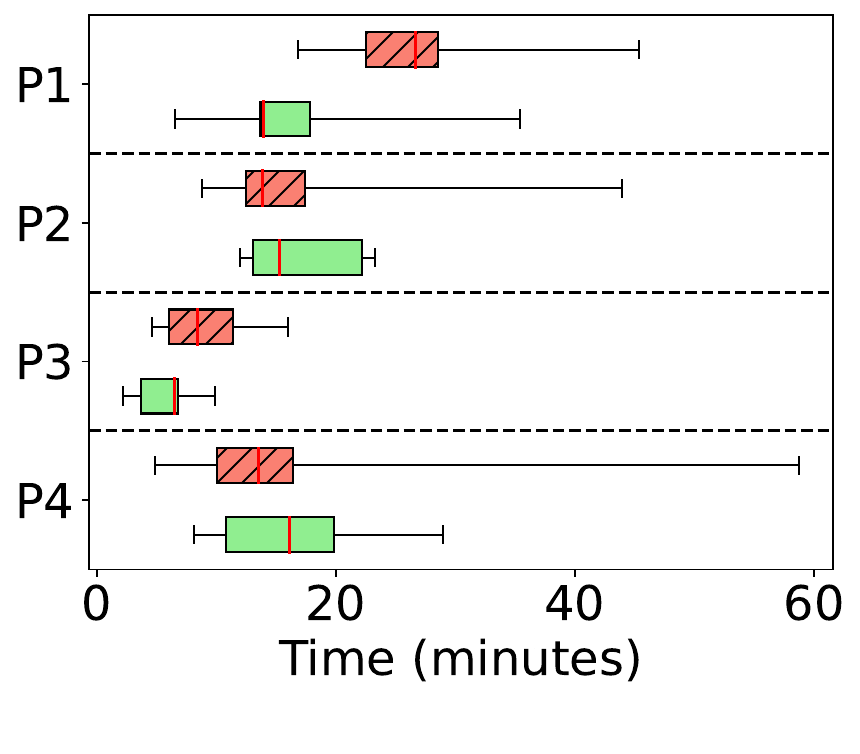} 
  }
  \hfill
  \subfloat[\label{fig:user-study:explanations:accuracy}]{
    \includegraphics[width=.55\textwidth]{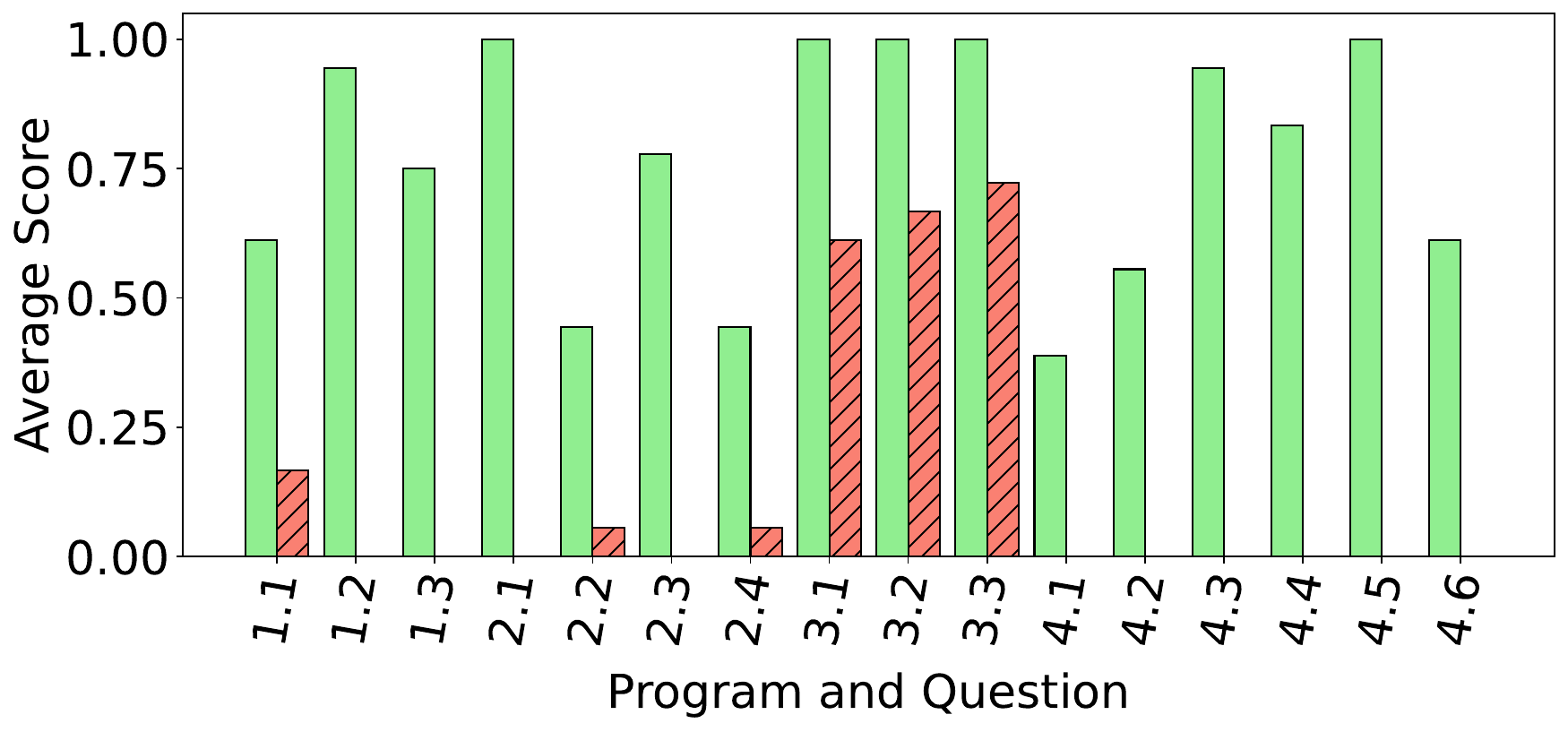} 
  }
  \hfill{}
  \includegraphics[width=.3\textwidth]{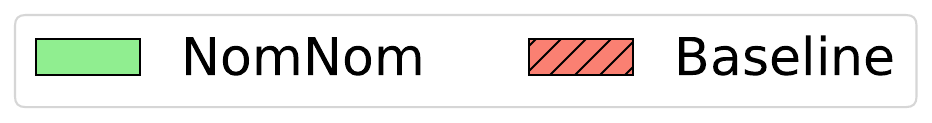}
\caption{Time needed and accuracy of participant responses when explaining how implementations
  worked (Study~1). The missing bars for the baseline correspond to questions that no participant
  was able to successfully answer.}
\label{fig:user-study:explanations}
\end{figure*}

For the first class of questions (i.e., what each subroutine did), we awarded responses with grades
of either 0, 0.5, or 1, indicating whether it adequately captured the high-level goal of the
function. For the second class of questions (i.e., walking us through an execution), we assessed
whether participants accurately described how each function made use of the auxiliary functions that
it called. For each such question, we normalized their scores from 0 to 1.
We measured the time needed by participants and their average scores. See Figure~%
\ref{fig:user-study:explanations}.

\begin{figure}
\centering
\includegraphics[width=.6\columnwidth]{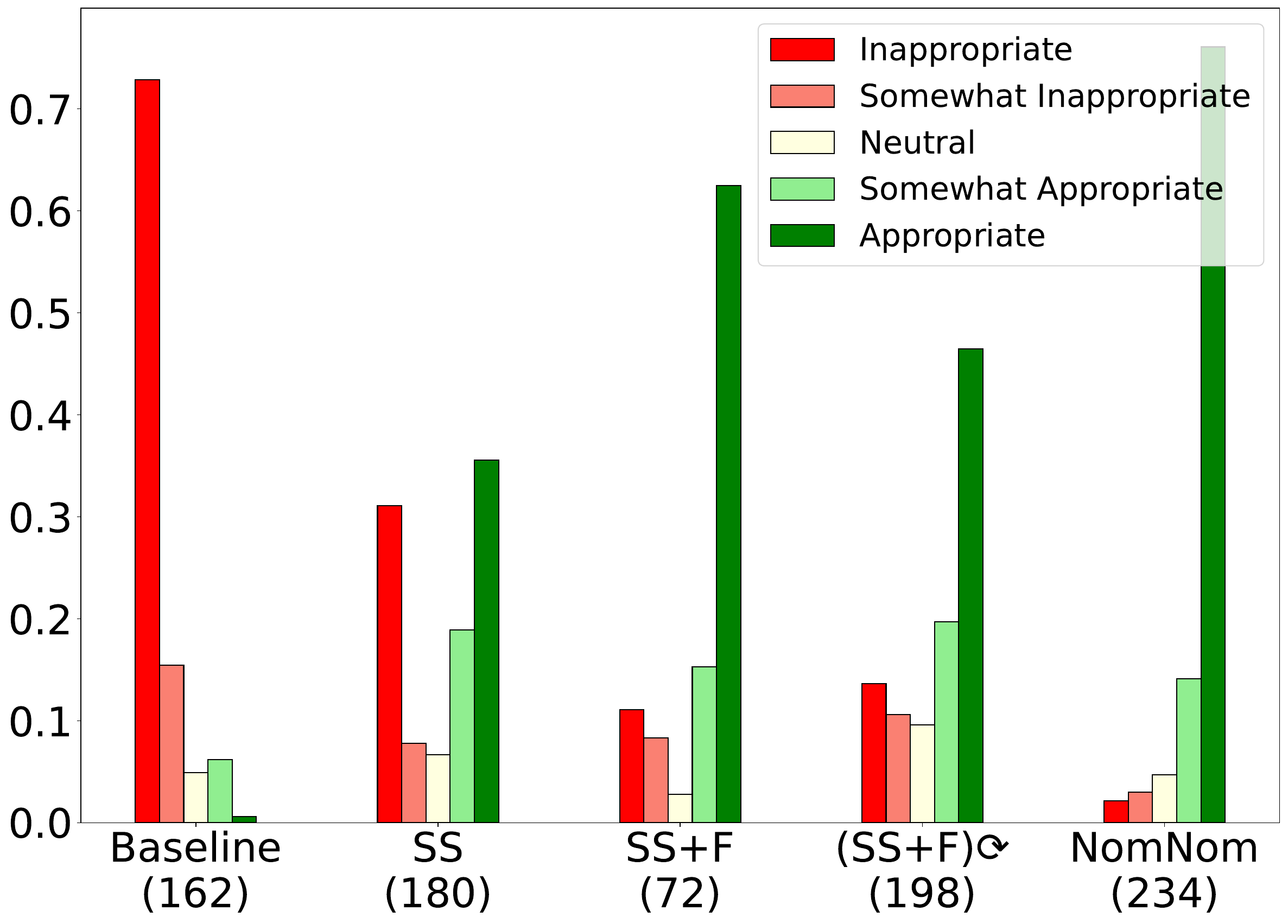}
\caption{Distribution of participant preferences among names suggested by different algorithms
  (Study~2). The numbers in parentheses indicate the total number of user responses collected for
  the corresponding naming algorithm.}
\label{fig:user-study:prefs}
\end{figure}

\paragraph{Study~2: Preferences.}
The second study consisted of four tasks in which we presented users with a program and asked them
to rate their preferences among different suggestions for function names on a five-point Likert
scale (``\emph{Inappropriate}'', ``\emph{somewhat inappropriate}'', ``\emph{neutral}'',
``\emph{somewhat appropriate}'', ``\emph{appropriate}''). We used the same programs as in the
previous study. All participants undertook this study in the same condition, and had access to local
subspecifications for each subroutine. We measure the distribution of their responses in Figure~%
\ref{fig:user-study:prefs}.

\mclearpage
\subsection{\ref{enu:user-study:purpose}: Understanding What Functions Do}
\label{sub:user-study:purpose}

We first measured the impact of the proposed function names on users' understanding of
the top-level purpose of each function, and the effect of the different naming algorithms on this
understanding. From Figure~\ref{fig:user-study:explanations:accuracy}, we observe that names
suggested by our tool unambiguously help users in determining the purpose of each function.

Notice that for a majority of questions, all responses from participants looking at names from the
baseline LLM were incorrect. We discovered that for the 12 subroutines in question, the baseline
algorithm never suggested an appropriate name: four of its suggestions were nonsensical, and eight
name suggestions were actually misleading.

By our estimate, all participants critically examined the programs that we showed them. When faced
with misleading names, they responded in a few different ways: one subset of participants chose to
skip the question, another group of participants sensed a mismatch but chose to trust the stated
names anyway, while the last subset disregarded the stated names upon discovering their
inappropriateness and attempted to manually recover the function specification. In any case, apart
from the program in Task~3, they were uniformly unable to discover the true purpose of the
corresponding functions.

In contrast, we judged that all names produced by the final algorithm were appropriate, leading
to massively higher accuracies when participants encountered questions in this condition.


\mclearpage
\subsection{\ref{enu:user-study:relation}: Understanding How They Work}
\label{sub:user-study:relation}

Our next question involved determining the effect of names on users' understanding of how functions
work. In particular, we asked users to walk us through an execution of the program, and focused on
whether they were able to articulate the relationships among the various subroutines, i.e., why and
how a particular subroutine called another.

Our measurements are contained in the last question for each task in Figure~%
\ref{fig:user-study:explanations:accuracy}. Similar to~\ref{enu:user-study:purpose}, we concluded
that participants have a much easier time understanding synthesized code when functions are
appropriately named. During the post-study debrief, participants reported being surprised by
``\emph{unnatural}'' implementations, and complained that the programs realize simple user
intentions in complicated ways. In addition, even when they had well-chosen function names, effort
was required to confirm their understanding and frame their responses to the questions being asked.
When they were unable to definitively understand the code, some participants chose to guess
functional relationships based on their names, while others opted to skip the question rather than
make tentative predictions.

Anecdotally, we also found a cascading effect in users' understanding as they went higher up in the
call graph towards the top-level functions: i.e., if they were unable to explain how a function
$f_2$ used an auxiliary function $f_3$ which was called in its body, then they were often also
unable to determine the working of a higher-level function $f_1$ which in turn called $f_2$.


\mclearpage
\subsection{\ref{enu:user-study:prefs}: Distribution of User Preferences}
\label{sub:user-study:prefs}

In our final research question, we attempted to determine how user preferences varied among names
produced by different naming algorithms. See Figure~\ref{fig:user-study:prefs}. From the figure, it
is clear that the progressive algorithmic improvements that we discussed in this paper result in
names that are well-liked by users. If we assign numerical values to these user preferences on a
0 (``\emph{Inappropriate}'')--1 (``\emph{Appropriate}'') scale, then the average score of names
produced by the baseline LLM is 0.11, while the average score of names produced by \Tool{} is 0.89.
Furthermore, the response rates show a similar trend as discussed in Section~\ref{sec:eval}, with
$\AlgPF$ producing the fewest number of suggestions.

Anecdotally, participants who undertook this study (Study~2) found it easier to indicate their
preferences, as compared to participants who had to provide more detailed accounts of how the
implementations worked (Study~1). Recall that we included the local input-output subspecifications
as part of this study: most participants made their judgments by simply checking for compatibility
between the proposed names and the subspecs.

%% file: src/limitations.tex
\section{Limitations} 
\label{sec:limitations}

We now discuss some limitations of our approach and the evaluation methodology employed in this
paper.


The first concern is whether function names are sufficient to help programmers understand the
synthesized implementation. How programmers choose identifier names~\cite{Feitelson:TSE2022,
Felienne:SCAM2017}, and their effect on program comprehension~\cite{ConciseAndConsistentNaming} has been the subject of
extensive research. Despite some research indicating otherwise~\cite{Beniamini:ICPC2017}, there is
broad agreement that good variable names are important~\cite{EffectsOfVariableNamesOnComprehension}. This is consistent with our
observations in Section~\ref{sec:user-study}, where programmers are able to better understand code
with appropriately chosen function names.

We might indeed have asked the language model to produce explanations in the form of longer
free-form comments: Note that our thesis in Section~\ref{sub:alg:filter} is that well-chosen
function names can be used to recover semantically equivalent implementations, and can thereby be
subject to experimental falsification. The experimental validation of free-form comments, whether
produced by a language model or a human programmer, is an important challenge for future work.


Another concern involves the brittleness of LLM outputs in response to minor changes in the provided
prompt. While this cannot be completely mitigated, we hope that the full text of prompts provided in
the supplementary material will at least partially address issues with reproducibility.

One might also object to human-written reference names being used for evaluation in Section~%
\ref{sec:eval}. To mitigate this concern, we had our reference names cross-verified by another
student programmer who is not among the authors of this paper.


A final concern involves potential biases in our choice of participants for the user study in
Section~\ref{sec:user-study}. Are these participants are representative of \emph{actual} users of
program synthesis tools? While we attempted to mitigate this concern by drawing broadly from
graduate and undergraduate students across engineering schools of various prominent American and
Canadian universities, conducting a larger study with working programmers is an important direction
of future work.

%% file: src/related.tex
\section{Related Work}
\label{sec:related}


\paragraph{Comprehension of Synthesized Programs}

While program synthesizers are meant to meet realize logical correctness specifications, their
output is rarely easy to understand. Synthesized programs are often obscure and complicated, with no
explanation on how the implementation works or meets the user intent. This obscurity partly stems
from the use of procedurally generated identifier names (\lstinline|v1|, \lstinline|v2|,
\lstinline|f1|, \lstinline|f2|, etc.) in synthesized code.

Human programmers use different practices to make code comprehensible, such as by writing detailed
comments~\cite{TheEffectOfModularization}, using meaningful names and identifiers~%
\cite{DescriptiveCompoundIdentifierNames}, or maintaining documentation of their rationale. Of these
approaches, the use of meaningful names has been found to make a significant contribution to improve
the comprehensibility~\cite{EffectsOfVariableNamesOnComprehension}. This is intuitive, as 70\% of
source code consists of identifiers~\cite{ConciseAndConsistentNaming}. Thus, the lack of meaningful
names in synthesized code inhibits understanding. Although the automatic generation of meaningful
function and variable names is a well-studied problem in software engineering: see~%
\cite{ASENameSurvey} for a survey, and Code2Vec~\cite{code2vec} and JSNice~\cite{JSNice} for
prominent examples. Still, the unintuitive nature of automatically generated programs makes it
challenging to apply existing techniques.

Research studies have looked into techniques to implement comprehensible code~%
\cite{Storey:ProgramComprehension}. Researchers have also discussed using various techniques to
help improve comprehension like program debugging~\cite{AlgDebug, DeltaDebug}, program slicing~%
\cite{1981:Slicing, KoMyers2004}, automatic summarization~\cite{AutomaticCodeSummarization}, and
user-guided program synthesis~\cite{2021:CHI:Glassman}. However, none of these techniques have been
used to generate names for synthesized programs. \cite{SS1} introduced subspecifications to
allow programmers to reason about individual parts of synthesized implementations. Our paper builds
on this idea and develops a technique to annotate implementations from program synthesizers with
meaningful identifier names.


\paragraph{Large Language Models in Program Synthesis}

Large language models (LLMs) have been shown to be surprisingly capable of generating code from
natural language specifications of programmer intent~\cite{Codex}. Such LLMs also have the potential
to improve the explainability of code by augmenting it with natural language explanations. However,
these large language models do not understand program semantics, and offer no guarantees about
quality and accuracy of the suggested code or explanations.

To provide guarantees with LLMs, researchers have suggested using LLMs as a complementary approach
to formal methods which can guarantee accuracy and adherence to specifications. For example,
Jigsaw~\cite{Jigsaw} is a program synthesis tool that augments LLMs with post-processing steps based
on program analysis and synthesis techniques, NLX~\cite{NLX} marries pre-trained natural language
models and component-based program synthesis for multi-modal program inference. In this paper, we
follow this approach of combining program synthesis and LLMs to augment synthesized programs with
explanatory names that help users understand the intent of the code.


\paragraph{Prompting}

The effective use of LLMs depends on carefully chosen prompts: Researchers have investigated various
prompting techniques like LLM programming~\cite{LMP} which is a combination of text prompting and
scripting, chain-of-thought prompting~\cite{ChainOfThoughtPrompting} which uses a series of
intermediate reasoning steps to perform complex reasoning, and probabilistic inference paradigm~%
\cite{ThinkSum} which probabilistically reasons over sets of objects using LLMs. While researchers
have investigated these approaches to improve the accuracy of language models, it is not evident how
to use these techniques to justify the appropriateness of names for program elements.

In this paper, we discuss two novel prompting approaches to improve the accuracy of explanatory
names generated by LLMs for synthesized programs. First, we expand the prompt using
subspecifications by recovering input-output data from the synthesis algorithm. Second, we conduct
sanity check of the algorithm by using another language model to validate the proposed explanations.

%% file: src/concl.tex
\section{Conclusion}
\label{sec:concl}

In this paper, we showed how to use a large language model (LLM) to annotate automatically
synthesized code with meaningful names.
Our procedure principally relies on a combination of two relatively simple techniques: \emph{first},
by including additional information about the implementation as part of the prompt, and \emph{next},
by validating the proposed names with an algorithmic sanity check. Both experiments and a user study
show the effectiveness of our technique in producing well-chosen function names.
Our research contributes to the emerging body of work on combining language models with formal
reasoning techniques. In future, we hope to extend these ideas to automatically produce free-form
comments, and also potentially techniques that confirm the validity of comments, identifiers and
other natural language artifacts in code.

%% file: src/app-prompts.tex
\section{Prompt Templates}
\label{app:prompts}

\begin{figure}
  \subfloat[\label{fig:user-study:template1}]{
      \frame{\includegraphics[width=.32\textwidth]{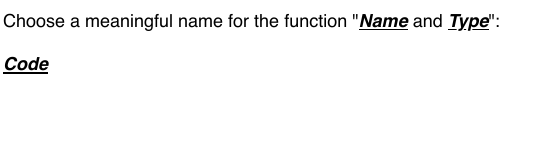}}
  }
  \hfill
  \subfloat[\label{fig:user-study:template2}]{
      \frame{\includegraphics[width=.32\textwidth]{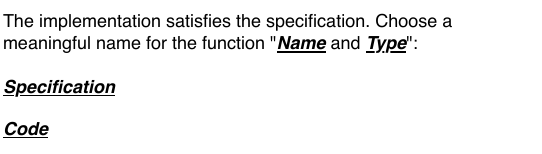}}
  }
  \hfill
  \subfloat[\label{fig:user-study:template3}]{
      \frame{\includegraphics[width=.32\textwidth]{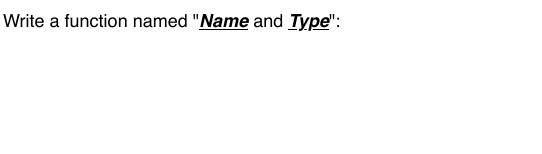}}
  }
\caption{Prompt templates utilized in the generation of names and code using the LLMs.
  Figure~\ref{fig:user-study:template1} is utilized to generate names for the baseline, while
  Figure~\ref{fig:user-study:template2} is employed to generate names for our algorithms.
  Additionally, Figure~\ref{fig:user-study:template3} is employed to generate programs for our
  filtering technique.}
\label{fig:prompts}
\end{figure}

Figure~\ref{fig:prompts} displays the prompt templates utilized in the generation of names and
code using the LLMs.

%% file: src/app-listings.tex
\section{Implementation Listings}
\label{app:listings}

We list the lambda-expressions emitted by DreamCoder in Figures~\ref{fig:listings:sort}--%
\ref{fig:listings:reverse} and our corresponding transliterated Python programs for all examples and
user study tasks in Figures~\ref{fig:listings:sortPython}--\ref{fig:listings:reversePython}. Note
that the lambda-expressions use de~Bruijn indices~\cite{deBruijn:1972, Schulman:deBruijn}, thereby
eliminating variable names.

\begin{figure}
\begin{lstlisting}
(lambda
  (map (lambda
         $\#$(lambda
           (lambda
             $\#$(lambda
               (car (filter $\textdollar$0 (lambda
                                (empty? (filter $\textdollar$1 (lambda
                                                     (gt? $\textdollar$0 $\textdollar$1))))))))
               (filter $\textdollar$1 (lambda
                          (gt? $\textdollar$1 (length (filter $\textdollar$2 (lambda
                                                     (gt? $\textdollar$1 $\textdollar$0)))))))))
       $\textdollar$1 (+ 1 $\textdollar$0)))
       (range (length $\textdollar$0)))
\end{lstlisting}
\caption{Sorting a list of numbers. The corresponding Python program is in
  Figure~\ref{fig:listings:sortPython}.}
\label{fig:listings:sort}
\end{figure}

\begin{figure}
\begin{lstlisting}
(lambda ($\#$(lambda (lambda ((lambda (gt? $\textdollar$0 0))
      ($\#$(lambda (lambda (length ($\#$(lambda (lambda ($\#$(lambda (lambda
          (fold $\textdollar$1 empty
                (lambda (lambda (fold ($\textdollar$2 $\textdollar$1 range) $\textdollar$0
                                      (lambda (lambda (cons $\textdollar$3 $\textdollar$2))))))))
          $\textdollar$1 (lambda (lambda (if ($\textdollar$2 $\textdollar$1) $\textdollar$3 empty)))))) $\textdollar$1
                (lambda (eq? $\textdollar$0 $\textdollar$1)))))) $\textdollar$0 $\textdollar$1))))
  (+ (+ (+ 1 (+ 1 1)) 1) 1)
  $\textdollar$0)
\end{lstlisting}
\caption{Checking if a list has 5. The corresponding Python program is in
  Figure~\ref{fig:listings:has5Python}.}
\label{fig:listings:has5}
\end{figure}

\begin{figure}
\begin{lstlisting}
(lambda (cdr (
    $\#$(lambda (lambda (map (lambda (index $\textdollar$0 $\textdollar$2)) (range $\textdollar$0))))
        (map (lambda $\textdollar$0) $\textdollar$0) (+ 1 (+ 1 1)))))
\end{lstlisting}
\caption{Getting second and third elements of a list. The corresponding Python program is in
  Figure~\ref{fig:listings:secondThirdElementsPython}.}
\label{fig:listings:secondThirdElements}
\end{figure}

\begin{figure}
\begin{lstlisting}
(lambda (fold $\textdollar$0 ($\#$(map (lambda (mod (+ $\textdollar$0 1) (+ 1 (+ 1 1))))) empty)
    (lambda (lambda ($\#$(lambda (lambda (fold $\textdollar$1 (cons $\textdollar$0 empty)
        (lambda (lambda (cons $\textdollar$1 $\textdollar$0)))))) $\textdollar$0 $\textdollar$1))))
\end{lstlisting}
\caption{Reversing a list. The corresponding Python program is in
  Figure~\ref{fig:listings:reversePython}.}
\label{fig:listings:reverse}
\end{figure}

\begin{figure}
\begin{lstlisting}
def a3(x3):
    def a31(x31):
        def a32(x32):
            return x31 < x32
        return len(list(filter(a32, x3)))==0
    return list(filter(a31, x3))[0]

def a2(x2):
    def a21(x21):
        def a22(x22):
            def a23(x23):
                return x22>x23
            return x21>len(list(filter(a23, x2)))
        return a3(list(filter(a22, x2)))
    return a21

def a1(x1):
    def a11(x11):
        return a2(x1)(x11+1)
    return list(map(a11, range(len(x1))))
\end{lstlisting}
\caption{Sorting a list of numbers.}
\label{fig:listings:sortPython}
\end{figure}

\begin{figure}
\begin{lstlisting}
import functools

def a5(x5):
    def a51(x51):
        def a52(x52, x53):
            def a54(x54, x55):
                return [x53]+x52
            return functools.reduce(a54, x51(x53)(range)[::-1], x52)
        return functools.reduce(a52, x5[::-1], [])
    return a51

def a4(x4):
    def a41(x41):
        def a42(x42):
            def a43(x43):
                if x41(x42):
                    return x4
                else:
                    return []
            return a43
        return a5(x4)(a42)
    return a41

def a3(x3):
    def a31(x31):
        def a32(x32):
            return x32==x31
        return len(a4(x3)(a32))
    return a31

def a2(x2):
    def a21(x21):
        def a22(x22):
            return x22>0
        return a22(a3(x2)(x21))
    return a21

def a1(x1):
    return a2(x1)(((1+(1+1))+1)+1)
\end{lstlisting}
\caption{Checking if a list has 5.}
\label{fig:listings:has5Python}
\end{figure}

\begin{figure}
\begin{lstlisting}
def a2(x2):
    def a21(x21):
        def a22(x22):
            return x2[x22]
        return list(map(a22, range(x21)))
    return a21

def a1(x1):
    def a11(x11):
        return x11
    return a2(a11(x1))(1+(1+1))[1:]
\end{lstlisting}
\caption{Getting second and third elements of a list.}
\label{fig:listings:secondThirdElementsPython}
\end{figure}

\begin{figure}
\begin{lstlisting}
import functools

def a3(x3):
    def a31(x31):
        def a32(x32, x33):
            return [x33] + x32
        return functools.reduce(a32, x3[::-1], [x31])
    return a31

def a2(x2):
    return (x2+1)%(1+(1+1))

def a1(x1):
    def a11(x11, x12):
        return a3(x11)(x12)
    return functools.reduce(a11, x1[::-1], list(map(a2, [])))
\end{lstlisting}
\caption{Reversing a list.}
\label{fig:listings:reversePython}
\end{figure}

\begin{figure}
\begin{lstlisting}
  def genNextGreaterValue(x):
    def f(i: int):
        greater = []
        for e in x:
            if e > i:
                greater.append(e)
        return min(greater)
    return f
\end{lstlisting}
\caption{Generated code by LLM for the generated name \lstinline|genNextGreaterValue|.}
\label{fig:listings:genNextGreaterValue}
\end{figure}

%% file: src/app-tasks.tex
\section{User Study Tasks}
\label{app:user-study}
Figure~\ref{table:user-study:names} includes all names generated by all algorithms used in two user
studies.

\begin{table}
\caption{The generated names used in two user studies. In the first user study, we only used names
  generated by the baseline and \Tool. In the second user study, we asked participants to rate the
  names generated by the baseline, $\AlgP$, $\AlgPF$, $\AlgPFR$, and \Tool. We omitted the names
  produced by Code2Vec because they were uniformly uninformative.}
\label{table:user-study:names}
\centering
\tiny
\scalebox{0.96}{\input{src/images/user-study/names.tex}}
\end{table}

%% file: src/images/user-study/names.tex
\begin{tabular}{>{\centering}m{0.05\textwidth}%
                >{\centering}m{0.05\textwidth}%
                >{\centering}m{0.14\textwidth}%
                >{\centering}m{0.14\textwidth}%
                >{\centering}m{0.14\textwidth}%
                >{\centering}m{0.16\textwidth}%
                >{\centering}m{0.16\textwidth}}
\toprule
  \textbf{Function} & \textbf{Code2Vec} & \textbf{Baseline}  & \textbf{SS}  & \textbf{SS+F}  & \textbf{(SS+F)$\lcirclearrowright$} & \textbf{NomNom} \tabularnewline
\midrule
  \textbf{T1.a3} & a3 & foldFuncComposer &  &  &  & appendListFunc \tabularnewline
  \textbf{T1.a2} & a2 & addModuloThree & getScaledX21Value & modulo$\_$plus$\_$one & modulo$\_$plus$\_$one & addOneModThree \tabularnewline
  \textbf{T1.a1} & a1 & & reverseOrderList & reverseList & reverseMapping & reverseList \tabularnewline
  \textbf{T2.a3} & a3 & getLowestUnusedValue & getMaxElement &  & getMaxElement & getLargestElement \tabularnewline
  \textbf{T2.a2} & a2 & getHighestValueInList & getNthElementInList &  &  & getOrderedElement \tabularnewline
  \textbf{T2.a1} & a1 & findMaxElementIndex & findMaxElementIndex &  & sortList & sortList \tabularnewline
  \textbf{T3.a2} & a2 & transformList &  &  & getFirstXElements & getFirstNItems \tabularnewline
  \textbf{T3.a1} & a1 &  &  &  & extractNextTwoElements & getSecondAndThirdElements \tabularnewline
  \textbf{T4.a5} & a5 &  & filterFives &  & filterAppendFunc & filterFunc \tabularnewline
  \textbf{T4.a4} & a4 &  & filterIntListByPredicate & filterListByPredicate & filterList & filterListByPredicate \tabularnewline
  \textbf{T4.a3} & a3 & summationFunc & countOccurrences &  & countElementsMatchingValue & countOccurrences \tabularnewline
  \textbf{T4.a2} & a2 & filterPositiveNumbers & hasElementsGreaterThan & containsNumber & containsNum & hasAtLeastOneOccurrence \tabularnewline
  \textbf{T4.a1} & a1 & isOddIntList & isEqual5s &  & checkForFiveInSequences & hasFiveInList \tabularnewline
\bottomrule
\end{tabular}

%% file: main.bbl

\begin{thebibliography}{64}


\ifx \showCODEN    \undefined \def \showCODEN     #1{\unskip}     \fi
\ifx \showDOI      \undefined \def \showDOI       #1{#1}\fi
\ifx \showISBNx    \undefined \def \showISBNx     #1{\unskip}     \fi
\ifx \showISBNxiii \undefined \def \showISBNxiii  #1{\unskip}     \fi
\ifx \showISSN     \undefined \def \showISSN      #1{\unskip}     \fi
\ifx \showLCCN     \undefined \def \showLCCN      #1{\unskip}     \fi
\ifx \shownote     \undefined \def \shownote      #1{#1}          \fi
\ifx \showarticletitle \undefined \def \showarticletitle #1{#1}   \fi
\ifx \showURL      \undefined \def \showURL       {\relax}        \fi
\providecommand\bibfield[2]{#2}
\providecommand\bibinfo[2]{#2}
\providecommand\natexlab[1]{#1}
\providecommand\showeprint[2][]{arXiv:#2}

\bibitem[Cop(2021)]%
        {Copilot}
 \bibinfo{year}{2021}\natexlab{}.
\newblock \bibinfo{title}{GitHub Copilot}.
\newblock \bibinfo{howpublished}{\url{https://copilot.github.com/}}.
\newblock


\bibitem[Allamanis et~al\mbox{.}(2015)]%
        {Allamanis:FSE2015}
\bibfield{author}{\bibinfo{person}{Miltiadis Allamanis}, \bibinfo{person}{Earl
  Barr}, \bibinfo{person}{Christian Bird}, {and} \bibinfo{person}{Charles
  Sutton}.} \bibinfo{year}{2015}\natexlab{}.
\newblock \showarticletitle{Suggesting Accurate Method and Class Names}. In
  \bibinfo{booktitle}{\emph{Proceedings of the 2015 10th Joint Meeting on
  Foundations of Software Engineering}} \emph{(\bibinfo{series}{ESEC/FSE
  2015})}. \bibinfo{publisher}{ACM}, \bibinfo{pages}{38--49}.
\newblock
\showISBNx{9781450336758}
\urldef\tempurl%
\url{https://doi.org/10.1145/2786805.2786849}
\showDOI{\tempurl}


\bibitem[Alon et~al\mbox{.}(2019)]%
        {code2vec}
\bibfield{author}{\bibinfo{person}{Uri Alon}, \bibinfo{person}{Meital
  Zilberstein}, \bibinfo{person}{Omer Levy}, {and} \bibinfo{person}{Eran
  Yahav}.} \bibinfo{year}{2019}\natexlab{}.
\newblock \showarticletitle{Code2vec: {L}earning Distributed Representations of
  Code}.
\newblock \bibinfo{journal}{\emph{Proceedings of the ACM on Programming
  Languages}} \bibinfo{volume}{3}, \bibinfo{number}{POPL}, Article
  \bibinfo{articleno}{40} (\bibinfo{date}{jan} \bibinfo{year}{2019}),
  \bibinfo{numpages}{29}~pages.
\newblock
\urldef\tempurl%
\url{https://doi.org/10.1145/3290353}
\showDOI{\tempurl}


\bibitem[Alur et~al\mbox{.}(2017)]%
        {EUSolver}
\bibfield{author}{\bibinfo{person}{Rajeev Alur}, \bibinfo{person}{Arjun
  Radhakrishna}, {and} \bibinfo{person}{Abhishek Udupa}.}
  \bibinfo{year}{2017}\natexlab{}.
\newblock \showarticletitle{Scaling Enumerative Program Synthesis via Divide
  and Conquer}. In \bibinfo{booktitle}{\emph{23rd International Conference on
  Tools and Algorithms for the Construction and Analysis of Systems (TACAS)}}
  \emph{(\bibinfo{series}{Lecture Notes in Computer Science},
  Vol.~\bibinfo{volume}{10205})}. \bibinfo{pages}{319--336}.
\newblock


\bibitem[Alur et~al\mbox{.}(2018)]%
        {SearchBasedProgramSynthesis}
\bibfield{author}{\bibinfo{person}{Rajeev Alur}, \bibinfo{person}{Rishabh
  Singh}, \bibinfo{person}{Dana Fisman}, {and} \bibinfo{person}{Armando
  Solar-Lezama}.} \bibinfo{year}{2018}\natexlab{}.
\newblock \showarticletitle{Search-Based Program Synthesis}.
\newblock \bibinfo{journal}{\emph{Commun. ACM}} \bibinfo{volume}{61},
  \bibinfo{number}{12} (\bibinfo{date}{nov} \bibinfo{year}{2018}),
  \bibinfo{pages}{84–93}.
\newblock
\showISSN{0001-0782}
\urldef\tempurl%
\url{https://doi.org/10.1145/3208071}
\showDOI{\tempurl}


\bibitem[An et~al\mbox{.}(2019)]%
        {2020R2}
\bibfield{author}{\bibinfo{person}{Shengwei An}, \bibinfo{person}{Rishabh
  Singh}, \bibinfo{person}{Sasa Misailovic}, {and} \bibinfo{person}{Roopsha
  Samanta}.} \bibinfo{year}{2019}\natexlab{}.
\newblock \showarticletitle{Augmented Example-Based Synthesis Using Relational
  Perturbation Properties}.
\newblock \bibinfo{journal}{\emph{Proceedings of the ACM on Programming
  Languages}} \bibinfo{volume}{4}, \bibinfo{number}{POPL}, Article
  \bibinfo{articleno}{56} (\bibinfo{date}{Dec.} \bibinfo{year}{2019}),
  \bibinfo{numpages}{24}~pages.
\newblock


\bibitem[Antoni et~al\mbox{.}(2016)]%
        {Qlose}
\bibfield{author}{\bibinfo{person}{Loris Antoni}, \bibinfo{person}{Roopsha
  Samanta}, {and} \bibinfo{person}{Rishabh Singh}.}
  \bibinfo{year}{2016}\natexlab{}.
\newblock \showarticletitle{Qlose: {P}rogram Repair with Quantiative
  Objectives}. In \bibinfo{booktitle}{\emph{27th International Conference on
  Computer Aided Verification}} \emph{(\bibinfo{series}{CAV})}.
\newblock


\bibitem[Avidan and Feitelson(2017)]%
        {EffectsOfVariableNamesOnComprehension}
\bibfield{author}{\bibinfo{person}{Eran Avidan} {and} \bibinfo{person}{Dror~G.
  Feitelson}.} \bibinfo{year}{2017}\natexlab{}.
\newblock \showarticletitle{Effects of Variable Names on Comprehension: An
  Empirical Study}. In \bibinfo{booktitle}{\emph{2017 IEEE/ACM 25th
  International Conference on Program Comprehension (ICPC)}}.
  \bibinfo{pages}{55--65}.
\newblock
\urldef\tempurl%
\url{https://doi.org/10.1109/ICPC.2017.27}
\showDOI{\tempurl}


\bibitem[Beniamini et~al\mbox{.}(2017)]%
        {Beniamini:ICPC2017}
\bibfield{author}{\bibinfo{person}{Gal Beniamini}, \bibinfo{person}{Sarah
  Gingichashvili}, \bibinfo{person}{Alon~Klein Orbach}, {and}
  \bibinfo{person}{Dror~G. Feitelson}.} \bibinfo{year}{2017}\natexlab{}.
\newblock \showarticletitle{Meaningful Identifier Names: The Case of
  Single-Letter Variables}. In \bibinfo{booktitle}{\emph{Proceedings of the
  25th International Conference on Program Comprehension}} (Buenos Aires,
  Argentina) \emph{(\bibinfo{series}{ICPC '17})}. \bibinfo{publisher}{IEEE
  Press}, \bibinfo{pages}{45–54}.
\newblock
\showISBNx{9781538605356}
\urldef\tempurl%
\url{https://doi.org/10.1109/ICPC.2017.18}
\showDOI{\tempurl}


\bibitem[Beurer-Kellner et~al\mbox{.}(2023)]%
        {LMP}
\bibfield{author}{\bibinfo{person}{Luca Beurer-Kellner}, \bibinfo{person}{Marc
  Fischer}, {and} \bibinfo{person}{Martin Vechev}.}
  \bibinfo{year}{2023}\natexlab{}.
\newblock \showarticletitle{Prompting Is Programming: {A} Query Language for
  Large Language Models}.
\newblock  \bibinfo{volume}{7}, \bibinfo{number}{PLDI}, Article
  \bibinfo{articleno}{186} (\bibinfo{year}{2023}),
  \bibinfo{numpages}{24}~pages.
\newblock


\bibitem[Caballero et~al\mbox{.}(2017)]%
        {AlgDebug}
\bibfield{author}{\bibinfo{person}{Rafael Caballero},
  \bibinfo{person}{Adri{\'a}n Riesco}, {and} \bibinfo{person}{Josep Silva}.}
  \bibinfo{year}{2017}\natexlab{}.
\newblock \showarticletitle{A Survey of Algorithmic Debugging}.
\newblock \bibinfo{journal}{\emph{Comput. Surveys}}  \bibinfo{volume}{50}
  (\bibinfo{date}{08} \bibinfo{year}{2017}), \bibinfo{pages}{1--35}.
\newblock


\bibitem[Cao et~al\mbox{.}(2023)]%
        {Babble}
\bibfield{author}{\bibinfo{person}{David Cao}, \bibinfo{person}{Rose Kunkel},
  \bibinfo{person}{Chandrakana Nandi}, \bibinfo{person}{Max Willsey},
  \bibinfo{person}{Zachary Tatlock}, {and} \bibinfo{person}{Nadia
  Polikarpova}.} \bibinfo{year}{2023}\natexlab{}.
\newblock \showarticletitle{Babble: {L}earning Better Abstractions with
  E-Graphs and Anti-Unification}.
\newblock \bibinfo{journal}{\emph{Proceedings of the ACM on Programming
  Languages}} \bibinfo{volume}{7}, \bibinfo{number}{POPL}, Article
  \bibinfo{articleno}{14} (\bibinfo{year}{2023}), \bibinfo{numpages}{29}~pages.
\newblock
\urldef\tempurl%
\url{https://doi.org/10.1145/3571207}
\showDOI{\tempurl}


\bibitem[Chen et~al\mbox{.}(2021)]%
        {Codex}
\bibfield{author}{\bibinfo{person}{Mark Chen}, \bibinfo{person}{Jerry Tworek},
  \bibinfo{person}{Heewoo Jun}, \bibinfo{person}{Qiming Yuan},
  \bibinfo{person}{Henrique~Ponde de Oliveira~Pinto}, \bibinfo{person}{Jared
  Kaplan}, \bibinfo{person}{Harri Edwards}, \bibinfo{person}{Yuri Burda},
  \bibinfo{person}{Nicholas Joseph}, \bibinfo{person}{Greg Brockman},
  \bibinfo{person}{Alex Ray}, \bibinfo{person}{Raul Puri},
  \bibinfo{person}{Gretchen Krueger}, \bibinfo{person}{Michael Petrov},
  \bibinfo{person}{Heidy Khlaaf}, \bibinfo{person}{Girish Sastry},
  \bibinfo{person}{Pamela Mishkin}, \bibinfo{person}{Brooke Chan},
  \bibinfo{person}{Scott Gray}, \bibinfo{person}{Nick Ryder},
  \bibinfo{person}{Mikhail Pavlov}, \bibinfo{person}{Alethea Power},
  \bibinfo{person}{Lukasz Kaiser}, \bibinfo{person}{Mohammad Bavarian},
  \bibinfo{person}{Clemens Winter}, \bibinfo{person}{Philippe Tillet},
  \bibinfo{person}{Felipe~Petroski Such}, \bibinfo{person}{Dave Cummings},
  \bibinfo{person}{Matthias Plappert}, \bibinfo{person}{Fotios Chantzis},
  \bibinfo{person}{Elizabeth Barnes}, \bibinfo{person}{Ariel Herbert-Voss},
  \bibinfo{person}{William~Hebgen Guss}, \bibinfo{person}{Alex Nichol},
  \bibinfo{person}{Alex Paino}, \bibinfo{person}{Nikolas Tezak},
  \bibinfo{person}{Jie Tang}, \bibinfo{person}{Igor Babuschkin},
  \bibinfo{person}{Suchir Balaji}, \bibinfo{person}{Shantanu Jain},
  \bibinfo{person}{William Saunders}, \bibinfo{person}{Christopher Hesse},
  \bibinfo{person}{Andrew~N. Carr}, \bibinfo{person}{Jan Leike},
  \bibinfo{person}{Josh Achiam}, \bibinfo{person}{Vedant Misra},
  \bibinfo{person}{Evan Morikawa}, \bibinfo{person}{Alec Radford},
  \bibinfo{person}{Matthew Knight}, \bibinfo{person}{Miles Brundage},
  \bibinfo{person}{Mira Murati}, \bibinfo{person}{Katie Mayer},
  \bibinfo{person}{Peter Welinder}, \bibinfo{person}{Bob McGrew},
  \bibinfo{person}{Dario Amodei}, \bibinfo{person}{Sam McCandlish},
  \bibinfo{person}{Ilya Sutskever}, {and} \bibinfo{person}{Wojciech Zaremba}.}
  \bibinfo{year}{2021}\natexlab{}.
\newblock \bibinfo{title}{Evaluating Large Language Models Trained on Code}.
\newblock
\newblock
\showeprint[arxiv]{2107.03374}~[cs.LG]


\bibitem[Cohen et~al\mbox{.}(2003)]%
        {comparisonOfStringMetrics}
\bibfield{author}{\bibinfo{person}{William Cohen}, \bibinfo{person}{Pradeep
  Ravikumar}, {and} \bibinfo{person}{Stephen Fienberg}.}
  \bibinfo{year}{2003}\natexlab{}.
\newblock \showarticletitle{A comparison of string metrics for matching names
  and records}. In \bibinfo{booktitle}{\emph{Kdd workshop on data cleaning and
  object consolidation}}, Vol.~\bibinfo{volume}{3}. \bibinfo{pages}{73--78}.
\newblock


\bibitem[{de Bruijn}(1972)]%
        {deBruijn:1972}
\bibfield{author}{\bibinfo{person}{Nicolaas~Govert {de Bruijn}}.}
  \bibinfo{year}{1972}\natexlab{}.
\newblock \showarticletitle{Lambda calculus notation with nameless dummies, a
  tool for automatic formula manipulation, with application to the
  {C}hurch-{R}osser theorem}.
\newblock \bibinfo{journal}{\emph{Indagationes Mathematicae}}
  \bibinfo{volume}{75}, \bibinfo{number}{5} (\bibinfo{year}{1972}),
  \bibinfo{pages}{381--392}.
\newblock
\showISSN{1385-7258}


\bibitem[Deissenboeck and Pizka(2006)]%
        {ConciseAndConsistentNaming}
\bibfield{author}{\bibinfo{person}{Florian Deissenboeck} {and}
  \bibinfo{person}{Markus Pizka}.} \bibinfo{year}{2006}\natexlab{}.
\newblock \showarticletitle{Concise and consistent naming}.
\newblock \bibinfo{journal}{\emph{Software Quality Journal}}
  \bibinfo{volume}{14} (\bibinfo{year}{2006}), \bibinfo{pages}{261--282}.
\newblock


\bibitem[Ellis et~al\mbox{.}(2021a)]%
        {DreamCoder}
\bibfield{author}{\bibinfo{person}{Kevin Ellis}, \bibinfo{person}{Catherine
  Wong}, \bibinfo{person}{Maxwell Nye}, \bibinfo{person}{Mathias
  Sabl\'e-Meyer}, \bibinfo{person}{Lucas Morales}, \bibinfo{person}{Luke
  Hewitt}, \bibinfo{person}{Luc Cary}, \bibinfo{person}{Armando Solar-Lezama},
  {and} \bibinfo{person}{Joshua Tenenbaum}.} \bibinfo{year}{2021}\natexlab{a}.
\newblock \showarticletitle{DreamCoder: {B}ootstrapping Inductive Program
  Synthesis with Wake-Sleep Library Learning}. In
  \bibinfo{booktitle}{\emph{Proceedings of the 42nd ACM SIGPLAN International
  Conference on Programming Language Design and Implementation}}
  \emph{(\bibinfo{series}{PLDI 2021})}. \bibinfo{publisher}{ACM},
  \bibinfo{pages}{835–850}.
\newblock
\showISBNx{9781450383912}


\bibitem[Ellis et~al\mbox{.}(2021b)]%
        {DreamCoder:Artifact}
\bibfield{author}{\bibinfo{person}{Kevin Ellis}, \bibinfo{person}{Catherine
  Wong}, \bibinfo{person}{Maxwell Nye}, \bibinfo{person}{Mathias
  Sabl\'{e}-Meyer}, \bibinfo{person}{Lucas~Lucas Morales},
  \bibinfo{person}{Luke Hewitt}, \bibinfo{person}{Luc Cary},
  \bibinfo{person}{Armando Solar-Lezama}, {and} \bibinfo{person}{Joshua
  Tenenbaum}.} \bibinfo{year}{2021}\natexlab{b}.
\newblock \bibinfo{booktitle}{\emph{DreamCoder Software and Data}}.
\newblock
\urldef\tempurl%
\url{https://doi.org/10.1145/3410302}
\showURL{%
\tempurl}


\bibitem[Feitelson et~al\mbox{.}(2022)]%
        {Feitelson:TSE2022}
\bibfield{author}{\bibinfo{person}{Dror~G. Feitelson}, \bibinfo{person}{Ayelet
  Mizrahi}, \bibinfo{person}{Nofar Noy}, \bibinfo{person}{Aviad~Ben Shabat},
  \bibinfo{person}{Or Eliyahu}, {and} \bibinfo{person}{Roy Sheffer}.}
  \bibinfo{year}{2022}\natexlab{}.
\newblock \showarticletitle{How Developers Choose Names}.
\newblock \bibinfo{journal}{\emph{{IEEE} Trans. Software Eng.}}
  \bibinfo{volume}{48}, \bibinfo{number}{2} (\bibinfo{year}{2022}),
  \bibinfo{pages}{37--52}.
\newblock
\urldef\tempurl%
\url{https://doi.org/10.1109/TSE.2020.2976920}
\showDOI{\tempurl}


\bibitem[Feng et~al\mbox{.}(2018)]%
        {Dillig:ConflictDrivenLearning}
\bibfield{author}{\bibinfo{person}{Yu Feng}, \bibinfo{person}{Ruben Martins},
  \bibinfo{person}{Osbert Bastani}, {and} \bibinfo{person}{Isil Dillig}.}
  \bibinfo{year}{2018}\natexlab{}.
\newblock \showarticletitle{Program Synthesis Using Conflict-Driven Learning}.
  In \bibinfo{booktitle}{\emph{Proceedings of the 39th ACM SIGPLAN Conference
  on Programming Language Design and Implementation}}
  \emph{(\bibinfo{series}{PLDI 2018})}. \bibinfo{publisher}{ACM},
  \bibinfo{pages}{420–435}.
\newblock
\showISBNx{9781450356985}


\bibitem[Feser et~al\mbox{.}(2015)]%
        {Feser:PLDI2015}
\bibfield{author}{\bibinfo{person}{John~K. Feser}, \bibinfo{person}{Swarat
  Chaudhuri}, {and} \bibinfo{person}{Isil Dillig}.}
  \bibinfo{year}{2015}\natexlab{}.
\newblock \showarticletitle{Synthesizing Data Structure Transformations from
  Input-Output Examples}. In \bibinfo{booktitle}{\emph{Proceedings of the 36th
  ACM SIGPLAN Conference on Programming Language Design and Implementation}}
  (Portland, OR, USA) \emph{(\bibinfo{series}{PLDI '15})}.
  \bibinfo{publisher}{Association for Computing Machinery},
  \bibinfo{address}{New York, NY, USA}, \bibinfo{pages}{229–239}.
\newblock
\showISBNx{9781450334686}
\urldef\tempurl%
\url{https://doi.org/10.1145/2737924.2737977}
\showDOI{\tempurl}


\bibitem[Gulwani(2011)]%
        {FlashFill}
\bibfield{author}{\bibinfo{person}{Sumit Gulwani}.}
  \bibinfo{year}{2011}\natexlab{}.
\newblock \showarticletitle{Automating String Processing in Spreadsheets Using
  Input-Output Examples}. In \bibinfo{booktitle}{\emph{Proceedings of the 38th
  Annual ACM SIGPLAN-SIGACT Symposium on Principles of Programming Languages}}
  \emph{(\bibinfo{series}{POPL})}. \bibinfo{publisher}{ACM},
  \bibinfo{pages}{317–330}.
\newblock
\showISBNx{9781450304900}


\bibitem[Gulwani et~al\mbox{.}(2017)]%
        {GulwaniPolozovRishabh:SynthesisSurvey}
\bibfield{author}{\bibinfo{person}{Sumit Gulwani}, \bibinfo{person}{Alex
  Polozov}, {and} \bibinfo{person}{Rishabh Singh}.}
  \bibinfo{year}{2017}\natexlab{}.
\newblock \bibinfo{booktitle}{\emph{Program Synthesis}}.
  Vol.~\bibinfo{volume}{4}.
\newblock \bibinfo{publisher}{NOW}. 1--119 pages.
\newblock
\urldef\tempurl%
\url{https://www.microsoft.com/en-us/research/publication/program-synthesis/}
\showURL{%
\tempurl}


\bibitem[Guo et~al\mbox{.}(2019)]%
        {Nadia:POPL2019}
\bibfield{author}{\bibinfo{person}{Zheng Guo}, \bibinfo{person}{Michael James},
  \bibinfo{person}{David Justo}, \bibinfo{person}{Jiaxiao Zhou},
  \bibinfo{person}{Ziteng Wang}, \bibinfo{person}{Ranjit Jhala}, {and}
  \bibinfo{person}{Nadia Polikarpova}.} \bibinfo{year}{2019}\natexlab{}.
\newblock \showarticletitle{Program Synthesis by Type-Guided Abstraction
  Refinement}.
\newblock \bibinfo{journal}{\emph{Proceedings of the ACM on Programming
  Languages}} \bibinfo{volume}{4}, \bibinfo{number}{POPL}, Article
  \bibinfo{articleno}{12} (\bibinfo{date}{Dec.} \bibinfo{year}{2019}),
  \bibinfo{numpages}{28}~pages.
\newblock


\bibitem[Handa and Rinard(2020)]%
        {2020R3}
\bibfield{author}{\bibinfo{person}{Shivam Handa} {and} \bibinfo{person}{Martin
  Rinard}.} \bibinfo{year}{2020}\natexlab{}.
\newblock \showarticletitle{Inductive Program Synthesis over Noisy Data}. In
  \bibinfo{booktitle}{\emph{Proceedings of the 28th ACM Joint Meeting on
  European Software Engineering Conference and Symposium on the Foundations of
  Software Engineering}} \emph{(\bibinfo{series}{ESEC/FSE 2020})}.
  \bibinfo{publisher}{ACM}, \bibinfo{pages}{87--98}.
\newblock
\showISBNx{9781450370431}


\bibitem[Jain et~al\mbox{.}(2022)]%
        {Jigsaw}
\bibfield{author}{\bibinfo{person}{Naman Jain}, \bibinfo{person}{Skanda
  Vaidyanath}, \bibinfo{person}{Arun Iyer}, \bibinfo{person}{Nagarajan
  Natarajan}, \bibinfo{person}{Suresh Parthasarathy}, \bibinfo{person}{Sriram
  Rajamani}, {and} \bibinfo{person}{Rahul Sharma}.}
  \bibinfo{year}{2022}\natexlab{}.
\newblock \showarticletitle{Jigsaw: {L}arge Language Models Meet Program
  Synthesis}. In \bibinfo{booktitle}{\emph{Proceedings of the 44th
  International Conference on Software Engineering}}
  \emph{(\bibinfo{series}{ICSE})}. \bibinfo{publisher}{ACM},
  \bibinfo{pages}{1219--1231}.
\newblock
\showISBNx{9781450392211}


\bibitem[Jaro(1989)]%
        {advancesInRecordLinkageMethodology}
\bibfield{author}{\bibinfo{person}{Matthew~A. Jaro}.}
  \bibinfo{year}{1989}\natexlab{}.
\newblock \showarticletitle{Advances in Record-Linkage Methodology as Applied
  to Matching the 1985 Census of Tampa, Florida}.
\newblock \bibinfo{journal}{\emph{J. Amer. Statist. Assoc.}}
  \bibinfo{volume}{84}, \bibinfo{number}{406} (\bibinfo{year}{1989}),
  \bibinfo{pages}{414--420}.
\newblock
\urldef\tempurl%
\url{https://doi.org/10.1080/01621459.1989.10478785}
\showDOI{\tempurl}
\showeprint{https://www.tandfonline.com/doi/pdf/10.1080/01621459.1989.10478785}


\bibitem[Ji et~al\mbox{.}(2020)]%
        {2020R1}
\bibfield{author}{\bibinfo{person}{Ruyi Ji}, \bibinfo{person}{Yican Sun},
  \bibinfo{person}{Yingfei Xiong}, {and} \bibinfo{person}{Zhenjiang Hu}.}
  \bibinfo{year}{2020}\natexlab{}.
\newblock \showarticletitle{Guiding Dynamic Programing via Structural
  Probability for Accelerating Programming by Example}.
\newblock \bibinfo{journal}{\emph{Proceedings of the ACM on Programming
  Languages}} \bibinfo{volume}{4}, \bibinfo{number}{OOPSLA}, Article
  \bibinfo{articleno}{224} (\bibinfo{date}{Nov.} \bibinfo{year}{2020}),
  \bibinfo{numpages}{29}~pages.
\newblock


\bibitem[Jiang et~al\mbox{.}(2019)]%
        {ASENameSurvey}
\bibfield{author}{\bibinfo{person}{Lin Jiang}, \bibinfo{person}{Hui Liu}, {and}
  \bibinfo{person}{He Jiang}.} \bibinfo{year}{2019}\natexlab{}.
\newblock \showarticletitle{Machine Learning Based Recommendation of Method
  Names: {H}ow Far are We}. In \bibinfo{booktitle}{\emph{34th IEEE/ACM
  International Conference on Automated Software Engineering}}
  \emph{(\bibinfo{series}{ASE})}. \bibinfo{pages}{602--614}.
\newblock
\urldef\tempurl%
\url{https://doi.org/10.1109/ASE.2019.00062}
\showDOI{\tempurl}


\bibitem[Ko and Myers(2004)]%
        {KoMyers2004}
\bibfield{author}{\bibinfo{person}{Amy Ko} {and} \bibinfo{person}{Brad Myers}.}
  \bibinfo{year}{2004}\natexlab{}.
\newblock \showarticletitle{Designing the {W}hyline: {A} Debugging Interface
  for Asking Questions About Program Behavior}. In
  \bibinfo{booktitle}{\emph{Proceedings of the SIGCHI Conference on Human
  Factors in Computing Systems}} \emph{(\bibinfo{series}{CHI})}.
  \bibinfo{publisher}{ACM}, \bibinfo{pages}{151--158}.
\newblock
\showISBNx{1581137028}


\bibitem[Kupferman and Vardi(2003)]%
        {Kupferman:Vacuity}
\bibfield{author}{\bibinfo{person}{Orna Kupferman} {and} \bibinfo{person}{Moshe
  Vardi}.} \bibinfo{year}{2003}\natexlab{}.
\newblock \showarticletitle{Vacuity Detection in Temporal Model Checking}.
\newblock \bibinfo{journal}{\emph{International Journal on Software Tools for
  Technology Transfer}} \bibinfo{volume}{4}, \bibinfo{number}{2}
  (\bibinfo{date}{Feb.} \bibinfo{year}{2003}), \bibinfo{pages}{224--233}.
\newblock
\showISSN{1433-2779}


\bibitem[Laich et~al\mbox{.}(2020)]%
        {2020R4}
\bibfield{author}{\bibinfo{person}{Larissa Laich}, \bibinfo{person}{Pavol
  Bielik}, {and} \bibinfo{person}{Martin Vechev}.}
  \bibinfo{year}{2020}\natexlab{}.
\newblock \showarticletitle{Guiding Program Synthesis by Learning to Generate
  Examples}. In \bibinfo{booktitle}{\emph{International Conference on Learning
  Representations}}.
\newblock
\urldef\tempurl%
\url{https://openreview.net/forum?id=BJl07ySKvS}
\showURL{%
\tempurl}


\bibitem[Lau et~al\mbox{.}(2000)]%
        {VSA}
\bibfield{author}{\bibinfo{person}{Tessa Lau}, \bibinfo{person}{Pedro
  Domingos}, {and} \bibinfo{person}{Daniel Weld}.}
  \bibinfo{year}{2000}\natexlab{}.
\newblock \showarticletitle{Version Space Algebra and Its Application to
  Programming by Demonstration}. In \bibinfo{booktitle}{\emph{Proceedings of
  the Seventeenth International Conference on Machine Learning}}
  \emph{(\bibinfo{series}{ICML})}. \bibinfo{publisher}{Morgan Kaufmann
  Publishers Inc.}, \bibinfo{pages}{527--–534}.
\newblock
\showISBNx{1558607072}


\bibitem[Le and Gulwani(2014)]%
        {FlashExtract}
\bibfield{author}{\bibinfo{person}{Vu Le} {and} \bibinfo{person}{Sumit
  Gulwani}.} \bibinfo{year}{2014}\natexlab{}.
\newblock \showarticletitle{FlashExtract: {A} Framework for Data Extraction by
  Examples}. In \bibinfo{booktitle}{\emph{Proceedings of the 35th ACM SIGPLAN
  Conference on Programming Language Design and Implementation}}
  \emph{(\bibinfo{series}{PLDI})}. \bibinfo{publisher}{ACM},
  \bibinfo{pages}{542–--553}.
\newblock
\showISBNx{9781450327848}


\bibitem[Lee et~al\mbox{.}(2018)]%
        {Euphony}
\bibfield{author}{\bibinfo{person}{Woosuk Lee}, \bibinfo{person}{Kihong Heo},
  \bibinfo{person}{Rajeev Alur}, {and} \bibinfo{person}{Mayur Naik}.}
  \bibinfo{year}{2018}\natexlab{}.
\newblock \showarticletitle{Accelerating Search-Based Program Synthesis Using
  Learned Probabilistic Models}. In \bibinfo{booktitle}{\emph{Proceedings of
  the 39th ACM SIGPLAN Conference on Programming Language Design and
  Implementation}} \emph{(\bibinfo{series}{PLDI 2018})}.
  \bibinfo{publisher}{ACM}, \bibinfo{pages}{436–449}.
\newblock
\showISBNx{9781450356985}


\bibitem[McClurg et~al\mbox{.}(2017)]%
        {Pavol:CAV2017:NetworkSynthesis}
\bibfield{author}{\bibinfo{person}{Jedidiah McClurg}, \bibinfo{person}{Hossein
  Hojjat}, {and} \bibinfo{person}{Pavol \v{C}ern\'y}.}
  \bibinfo{year}{2017}\natexlab{}.
\newblock \showarticletitle{Synchronization Synthesis for Network Programs}. In
  \bibinfo{booktitle}{\emph{Computer Aided Verification}}.
  \bibinfo{publisher}{Springer}, \bibinfo{pages}{301--321}.
\newblock
\showISBNx{978-3-319-63390-9}


\bibitem[Miltner et~al\mbox{.}(2022)]%
        {Dillig:Burst}
\bibfield{author}{\bibinfo{person}{Anders Miltner},
  \bibinfo{person}{Adrian~Trejo Nu\~{n}ez}, \bibinfo{person}{Ana Brendel},
  \bibinfo{person}{Swarat Chaudhuri}, {and} \bibinfo{person}{Isil Dillig}.}
  \bibinfo{year}{2022}\natexlab{}.
\newblock \showarticletitle{Bottom-up Synthesis of Recursive Functional
  Programs Using Angelic Execution}.
\newblock \bibinfo{journal}{\emph{Proc. ACM Program. Lang.}}
  \bibinfo{volume}{6}, \bibinfo{number}{POPL}, Article \bibinfo{articleno}{21}
  (\bibinfo{date}{jan} \bibinfo{year}{2022}), \bibinfo{numpages}{29}~pages.
\newblock
\urldef\tempurl%
\url{https://doi.org/10.1145/3498682}
\showDOI{\tempurl}


\bibitem[Nazari et~al\mbox{.}(2023)]%
        {SS1}
\bibfield{author}{\bibinfo{person}{Amirmohammad Nazari}, \bibinfo{person}{Yifei
  Huang}, \bibinfo{person}{Roopsha Samanta}, \bibinfo{person}{Arjun
  Radhakrishna}, {and} \bibinfo{person}{Mukund Raghothaman}.}
  \bibinfo{year}{2023}\natexlab{}.
\newblock \showarticletitle{Explainable Program Synthesis by Localizing
  Specifications}.
\newblock \bibinfo{journal}{\emph{Proceedings of the ACM on Programming
  Languages}} \bibinfo{volume}{7}, \bibinfo{number}{OOPSLA2}, Article
  \bibinfo{articleno}{298} (\bibinfo{date}{oct} \bibinfo{year}{2023}),
  \bibinfo{numpages}{25}~pages.
\newblock
\urldef\tempurl%
\url{https://doi.org/10.1145/3622874}
\showDOI{\tempurl}


\bibitem[Nguyen et~al\mbox{.}(2013)]%
        {Semfix}
\bibfield{author}{\bibinfo{person}{Hoang Duong~Thien Nguyen},
  \bibinfo{person}{Dawei Qi}, \bibinfo{person}{Abhik Roychoudhury}, {and}
  \bibinfo{person}{Satish Chandra}.} \bibinfo{year}{2013}\natexlab{}.
\newblock \showarticletitle{SemFix: {P}rogram Repair via Semantic Analysis}. In
  \bibinfo{booktitle}{\emph{Proceedings of the 2013 International Conference on
  Software Engineering}} \emph{(\bibinfo{series}{ICSE})}.
  \bibinfo{publisher}{IEEE Press}, \bibinfo{pages}{772–781}.
\newblock
\showISBNx{9781467330763}


\bibitem[Nye et~al\mbox{.}(2021)]%
        {Scratchpads}
\bibfield{author}{\bibinfo{person}{Maxwell Nye}, \bibinfo{person}{Anders~Johan
  Andreassen}, \bibinfo{person}{Guy Gur-Ari}, \bibinfo{person}{Henryk
  Michalewski}, \bibinfo{person}{Jacob Austin}, \bibinfo{person}{David Bieber},
  \bibinfo{person}{David Dohan}, \bibinfo{person}{Aitor Lewkowycz},
  \bibinfo{person}{Maarten Bosma}, \bibinfo{person}{David Luan},
  \bibinfo{person}{Charles Sutton}, {and} \bibinfo{person}{Augustus Odena}.}
  \bibinfo{year}{2021}\natexlab{}.
\newblock \bibinfo{title}{Show Your Work: {S}cratchpads for Intermediate
  Computation with Language Models}.
\newblock
\newblock
\showeprint[arxiv]{2112.00114}~[cs.LG]


\bibitem[OpenAI(2023)]%
        {GPT4}
\bibfield{author}{\bibinfo{person}{OpenAI}.} \bibinfo{year}{2023}\natexlab{}.
\newblock \bibinfo{title}{GPT-4 Technical Report}.
\newblock
\newblock
\showeprint[arxiv]{2303.08774}~[cs.CL]


\bibitem[Osera and Zdancewic(2015)]%
        {PeterMichael:PLDI2015}
\bibfield{author}{\bibinfo{person}{Peter-Michael Osera} {and}
  \bibinfo{person}{Steve Zdancewic}.} \bibinfo{year}{2015}\natexlab{}.
\newblock \showarticletitle{Type-and-Example-Directed Program Synthesis}. In
  \bibinfo{booktitle}{\emph{Proceedings of the 36th ACM SIGPLAN Conference on
  Programming Language Design and Implementation}}
  \emph{(\bibinfo{series}{PLDI})}. \bibinfo{publisher}{ACM},
  \bibinfo{pages}{619--–630}.
\newblock
\showISBNx{9781450334686}


\bibitem[Ozturkler et~al\mbox{.}(2023)]%
        {ThinkSum}
\bibfield{author}{\bibinfo{person}{Batu Ozturkler}, \bibinfo{person}{Nikolay
  Malkin}, \bibinfo{person}{Zhen Wang}, {and} \bibinfo{person}{Nebojsa Jojic}.}
  \bibinfo{year}{2023}\natexlab{}.
\newblock \bibinfo{title}{ThinkSum: {P}robabilistic reasoning over sets using
  large language models}.
\newblock
\newblock
\showeprint[arxiv]{2210.01293}~[cs.CL]


\bibitem[Pal et~al\mbox{.}(2023)]%
        {Enumo}
\bibfield{author}{\bibinfo{person}{Anjali Pal}, \bibinfo{person}{Brett Saiki},
  \bibinfo{person}{Ryab Tjoa}, \bibinfo{person}{Cynthia Richey},
  \bibinfo{person}{Amy Zhu}, \bibinfo{person}{Oliver Flatt},
  \bibinfo{person}{Max Willsey}, \bibinfo{person}{Zachary Tatlock}, {and}
  \bibinfo{person}{Chandrakana Nandi}.} \bibinfo{year}{2023}\natexlab{}.
\newblock \showarticletitle{Equality Saturation Theory Exploration \'{a} La
  Carte}.
\newblock \bibinfo{journal}{\emph{Proceedings of the ACM on Programming
  Languages}} \bibinfo{volume}{7}, \bibinfo{number}{OOPSLA2}, Article
  \bibinfo{articleno}{258} (\bibinfo{year}{2023}),
  \bibinfo{numpages}{29}~pages.
\newblock
\urldef\tempurl%
\url{https://doi.org/10.1145/3622834}
\showDOI{\tempurl}


\bibitem[Pho\-thilimthana et~al\mbox{.}(2014)]%
        {Chlorophyll}
\bibfield{author}{\bibinfo{person}{Phitchaya~Mangpo Pho\-thilimthana},
  \bibinfo{person}{Tikhon Jelvis}, \bibinfo{person}{Rohin Shah},
  \bibinfo{person}{Nishant Totla}, \bibinfo{person}{Sarah Chasins}, {and}
  \bibinfo{person}{Rastislav Bodik}.} \bibinfo{year}{2014}\natexlab{}.
\newblock \showarticletitle{Chlorophyll: {S}ynthesis-Aided Compiler for
  Low-Power Spatial Architectures}. In \bibinfo{booktitle}{\emph{Proceedings of
  the 35th ACM SIGPLAN Conference on Programming Language Design and
  Implementation}} \emph{(\bibinfo{series}{PLDI})}. \bibinfo{publisher}{ACM},
  \bibinfo{pages}{396--–407}.
\newblock
\showISBNx{9781450327848}


\bibitem[Polikarpova et~al\mbox{.}(2016)]%
        {Polikarpova:PLDI2016}
\bibfield{author}{\bibinfo{person}{Nadia Polikarpova}, \bibinfo{person}{Ivan
  Kuraj}, {and} \bibinfo{person}{Armando Solar-Lezama}.}
  \bibinfo{year}{2016}\natexlab{}.
\newblock \showarticletitle{Program Synthesis from Polymorphic Refinement
  Types}. In \bibinfo{booktitle}{\emph{Proceedings of the 37th ACM SIGPLAN
  Conference on Programming Language Design and Implementation}} (Santa
  Barbara, CA, USA) \emph{(\bibinfo{series}{PLDI '16})}.
  \bibinfo{publisher}{Association for Computing Machinery},
  \bibinfo{address}{New York, NY, USA}, \bibinfo{pages}{522–538}.
\newblock
\showISBNx{9781450342612}
\urldef\tempurl%
\url{https://doi.org/10.1145/2908080.2908093}
\showDOI{\tempurl}


\bibitem[Rahmani et~al\mbox{.}(2021)]%
        {NLX}
\bibfield{author}{\bibinfo{person}{Kia Rahmani}, \bibinfo{person}{Mohammad
  Raza}, \bibinfo{person}{Sumit Gulwani}, \bibinfo{person}{Vu Le},
  \bibinfo{person}{Daniel Morris}, \bibinfo{person}{Arjun Radhakrishna},
  \bibinfo{person}{Gustavo Soares}, {and} \bibinfo{person}{Ashish Tiwari}.}
  \bibinfo{year}{2021}\natexlab{}.
\newblock \showarticletitle{Multi-Modal Program Inference: {A} Marriage of
  Pre-Trained Language Models and Component-Based Synthesis}.
\newblock \bibinfo{journal}{\emph{Proceedings of the ACM on Programming
  Languages}} \bibinfo{volume}{5}, \bibinfo{number}{OOPSLA}, Article
  \bibinfo{articleno}{158} (\bibinfo{date}{oct} \bibinfo{year}{2021}),
  \bibinfo{numpages}{29}~pages.
\newblock


\bibitem[Raychev et~al\mbox{.}(2015)]%
        {JSNice}
\bibfield{author}{\bibinfo{person}{Veselin Raychev}, \bibinfo{person}{Martin
  Vechev}, {and} \bibinfo{person}{Andreas Krause}.}
  \bibinfo{year}{2015}\natexlab{}.
\newblock \showarticletitle{Predicting Program Properties from "Big Code"}. In
  \bibinfo{booktitle}{\emph{Proceedings of the 42nd Annual ACM SIGPLAN-SIGACT
  Symposium on Principles of Programming Languages}}
  \emph{(\bibinfo{series}{POPL '15})}. \bibinfo{publisher}{ACM},
  \bibinfo{pages}{111–124}.
\newblock
\showISBNx{9781450333009}


\bibitem[Schankin et~al\mbox{.}(2018)]%
        {DescriptiveCompoundIdentifierNames}
\bibfield{author}{\bibinfo{person}{Andrea Schankin}, \bibinfo{person}{Annika
  Berger}, \bibinfo{person}{Daniel~V. Holt}, \bibinfo{person}{Johannes~C.
  Hofmeister}, \bibinfo{person}{Till Riedel}, {and} \bibinfo{person}{Michael
  Beigl}.} \bibinfo{year}{2018}\natexlab{}.
\newblock \showarticletitle{Descriptive Compound Identifier Names Improve
  Source Code Comprehension}. In \bibinfo{booktitle}{\emph{Proceedings of the
  26th Conference on Program Comprehension}} (Gothenburg, Sweden)
  \emph{(\bibinfo{series}{ICPC '18})}. \bibinfo{publisher}{Association for
  Computing Machinery}, \bibinfo{address}{New York, NY, USA},
  \bibinfo{pages}{31–40}.
\newblock
\showISBNx{9781450357142}
\urldef\tempurl%
\url{https://doi.org/10.1145/3196321.3196332}
\showDOI{\tempurl}


\bibitem[Schkufza et~al\mbox{.}(2013)]%
        {Stoke}
\bibfield{author}{\bibinfo{person}{Eric Schkufza}, \bibinfo{person}{Rahul
  Sharma}, {and} \bibinfo{person}{Alex Aiken}.}
  \bibinfo{year}{2013}\natexlab{}.
\newblock \showarticletitle{Stochastic Superoptimization}. In
  \bibinfo{booktitle}{\emph{Proceedings of the 18th International Conference on
  Architectural Support for Programming Languages and Operating Systems}}
  \emph{(\bibinfo{series}{ASPLOS})}. \bibinfo{publisher}{ACM},
  \bibinfo{pages}{305–316}.
\newblock
\showISBNx{9781450318709}


\bibitem[Shulman(2021)]%
        {Schulman:deBruijn}
\bibfield{author}{\bibinfo{person}{Mike Shulman}.}
  \bibinfo{year}{2021}\natexlab{}.
\newblock \bibinfo{title}{You Could Have Invented de~{B}ruijn Indices}.
\newblock
  \bibinfo{howpublished}{\url{https://golem.ph.utexas.edu/category/2021/08/you_could_have_invented_de_bru.html}}.
\newblock


\bibitem[Singh(2016)]%
        {Rishabh:BlinkFill}
\bibfield{author}{\bibinfo{person}{Rishabh Singh}.}
  \bibinfo{year}{2016}\natexlab{}.
\newblock \showarticletitle{BlinkFill: {S}emi-Supervised Programming by Example
  for Syntactic String Transformations}.
\newblock \bibinfo{journal}{\emph{Proceedings of the VLDB Endowment}}
  \bibinfo{volume}{9}, \bibinfo{number}{10} (\bibinfo{date}{June}
  \bibinfo{year}{2016}), \bibinfo{pages}{816–827}.
\newblock
\showISSN{2150-8097}


\bibitem[Singh et~al\mbox{.}(2013)]%
        {Rishabh:AutoTutor}
\bibfield{author}{\bibinfo{person}{Rishabh Singh}, \bibinfo{person}{Sumit
  Gulwani}, {and} \bibinfo{person}{Armando Solar-Lezama}.}
  \bibinfo{year}{2013}\natexlab{}.
\newblock \showarticletitle{Automated Feedback Generation for Introductory
  Programming Assignments}. In \bibinfo{booktitle}{\emph{Proceedings of the
  34th ACM SIGPLAN Conference on Programming Language Design and
  Implementation}} \emph{(\bibinfo{series}{PLDI})}. \bibinfo{publisher}{ACM},
  \bibinfo{pages}{15–26}.
\newblock
\showISBNx{9781450320146}


\bibitem[Storey(2005)]%
        {Storey:ProgramComprehension}
\bibfield{author}{\bibinfo{person}{M.-A. Storey}.}
  \bibinfo{year}{2005}\natexlab{}.
\newblock \showarticletitle{Theories, methods and tools in program
  comprehension: past, present and future}. In \bibinfo{booktitle}{\emph{13th
  International Workshop on Program Comprehension (IWPC'05)}}.
  \bibinfo{pages}{181--191}.
\newblock
\urldef\tempurl%
\url{https://doi.org/10.1109/WPC.2005.38}
\showDOI{\tempurl}


\bibitem[Swidan et~al\mbox{.}(2017)]%
        {Felienne:SCAM2017}
\bibfield{author}{\bibinfo{person}{Alaaeddin Swidan},
  \bibinfo{person}{Alexander Serebrenik}, {and} \bibinfo{person}{Felienne
  Hermans}.} \bibinfo{year}{2017}\natexlab{}.
\newblock \showarticletitle{How do Scratch Programmers Name Variables and
  Procedures?}. In \bibinfo{booktitle}{\emph{17th {IEEE} International Working
  Conference on Source Code Analysis and Manipulation, {SCAM} 2017, Shanghai,
  China, September 17-18, 2017}}. \bibinfo{publisher}{{IEEE} Computer Society},
  \bibinfo{pages}{51--60}.
\newblock
\urldef\tempurl%
\url{https://doi.org/10.1109/SCAM.2017.12}
\showDOI{\tempurl}


\bibitem[Touvron et~al\mbox{.}(2023)]%
        {LlaMA}
\bibfield{author}{\bibinfo{person}{Hugo Touvron}, \bibinfo{person}{Thibaut
  Lavril}, \bibinfo{person}{Gautier Izacard}, \bibinfo{person}{Xavier
  Martinet}, \bibinfo{person}{Marie-Anne Lachaux}, \bibinfo{person}{Timothée
  Lacroix}, \bibinfo{person}{Baptiste Rozière}, \bibinfo{person}{Naman Goyal},
  \bibinfo{person}{Eric Hambro}, \bibinfo{person}{Faisal Azhar},
  \bibinfo{person}{Aurelien Rodriguez}, \bibinfo{person}{Armand Joulin},
  \bibinfo{person}{Edouard Grave}, {and} \bibinfo{person}{Guillaume Lample}.}
  \bibinfo{year}{2023}\natexlab{}.
\newblock \bibinfo{title}{LLaMA: {O}pen and Efficient Foundation Language
  Models}.
\newblock
\newblock
\showeprint[arxiv]{2302.13971}~[cs.CL]


\bibitem[Udupa et~al\mbox{.}(2013)]%
        {Transit}
\bibfield{author}{\bibinfo{person}{Abhishek Udupa}, \bibinfo{person}{Arun
  Raghavan}, \bibinfo{person}{Jyotirmoy Deshmukh}, \bibinfo{person}{Sela
  Mador-Haim}, \bibinfo{person}{Milo Martin}, {and} \bibinfo{person}{Rajeev
  Alur}.} \bibinfo{year}{2013}\natexlab{}.
\newblock \showarticletitle{Transit: {S}pecifying Protocols with Concolic
  Snippets}. In \bibinfo{booktitle}{\emph{Proceedings of the 34th ACM SIGPLAN
  Conference on Programming Language Design and Implementation}}
  \emph{(\bibinfo{series}{PLDI})}. \bibinfo{publisher}{ACM},
  \bibinfo{pages}{287–296}.
\newblock
\showISBNx{9781450320146}


\bibitem[Wei et~al\mbox{.}(2022)]%
        {ChainOfThoughtPrompting}
\bibfield{author}{\bibinfo{person}{Jason Wei}, \bibinfo{person}{Xuezhi Wang},
  \bibinfo{person}{Dale Schuurmans}, \bibinfo{person}{Maarten Bosma},
  \bibinfo{person}{Ed~H. Chi}, \bibinfo{person}{Quoc Le}, {and}
  \bibinfo{person}{Denny Zhou}.} \bibinfo{year}{2022}\natexlab{}.
\newblock \showarticletitle{Chain of Thought Prompting Elicits Reasoning in
  Large Language Models}.
\newblock \bibinfo{journal}{\emph{CoRR}}  \bibinfo{volume}{abs/2201.11903}
  (\bibinfo{year}{2022}).
\newblock
\showeprint[arXiv]{2201.11903}


\bibitem[Weiser(1981)]%
        {1981:Slicing}
\bibfield{author}{\bibinfo{person}{Mark Weiser}.}
  \bibinfo{year}{1981}\natexlab{}.
\newblock \showarticletitle{Program Slicing}. In
  \bibinfo{booktitle}{\emph{Proceedings of the 5th International Conference on
  Software Engineering}} \emph{(\bibinfo{series}{ICSE})}.
  \bibinfo{publisher}{IEEE Press}, \bibinfo{pages}{439--449}.
\newblock
\showISBNx{0897911466}


\bibitem[Woodfield et~al\mbox{.}(1981)]%
        {TheEffectOfModularization}
\bibfield{author}{\bibinfo{person}{S.~N. Woodfield}, \bibinfo{person}{H.~E.
  Dunsmore}, {and} \bibinfo{person}{V.~Y. Shen}.}
  \bibinfo{year}{1981}\natexlab{}.
\newblock \showarticletitle{The Effect of Modularization and Comments on
  Program Comprehension}. In \bibinfo{booktitle}{\emph{Proceedings of the 5th
  International Conference on Software Engineering}} (San Diego, California,
  USA) \emph{(\bibinfo{series}{ICSE '81})}. \bibinfo{publisher}{IEEE Press},
  \bibinfo{pages}{215–223}.
\newblock
\showISBNx{0897911466}


\bibitem[Yuan et~al\mbox{.}(2023)]%
        {Roopsha:PLDI2023}
\bibfield{author}{\bibinfo{person}{Yongwei Yuan}, \bibinfo{person}{Arjun
  Radhakrishna}, {and} \bibinfo{person}{Roopsha Samanta}.}
  \bibinfo{year}{2023}\natexlab{}.
\newblock \showarticletitle{Trace-Guided Inductive Synthesis of Recursive
  Functional Programs}.
\newblock \bibinfo{journal}{\emph{Proceedings of the ACM on Programming
  Languages}} \bibinfo{volume}{7}, \bibinfo{number}{PLDI}, Article
  \bibinfo{articleno}{141} (\bibinfo{year}{2023}),
  \bibinfo{numpages}{24}~pages.
\newblock


\bibitem[Zeller(1999)]%
        {DeltaDebug}
\bibfield{author}{\bibinfo{person}{Andreas Zeller}.}
  \bibinfo{year}{1999}\natexlab{}.
\newblock \showarticletitle{Yesterday, My Program Worked. {T}oday, It Does Not.
  {W}hy?}. In \bibinfo{booktitle}{\emph{Proceedings of the 7th European
  Software Engineering Conference Held Jointly with the 7th ACM SIGSOFT
  International Symposium on Foundations of Software Engineering}}
  \emph{(\bibinfo{series}{ESEC/FSE})}. \bibinfo{publisher}{Springer},
  \bibinfo{pages}{253--267}.
\newblock
\showISBNx{3540665382}


\bibitem[Zhang et~al\mbox{.}(2021)]%
        {2021:CHI:Glassman}
\bibfield{author}{\bibinfo{person}{Tianyi Zhang}, \bibinfo{person}{Zhiyang
  Chen}, \bibinfo{person}{Yuanli Zhu}, \bibinfo{person}{Priyan Vaithilingam},
  \bibinfo{person}{Xinyu Wang}, {and} \bibinfo{person}{Elena Glassman}.}
  \bibinfo{year}{2021}\natexlab{}.
\newblock \showarticletitle{Interpretable Program Synthesis}. In
  \bibinfo{booktitle}{\emph{Proceedings of the 2021 CHI Conference on Human
  Factors in Computing Systems}}. \bibinfo{publisher}{ACM},
  \bibinfo{numpages}{16}~pages.
\newblock
\showISBNx{9781450380966}


\bibitem[Zhu and Pan(2019)]%
        {AutomaticCodeSummarization}
\bibfield{author}{\bibinfo{person}{Yuxiang Zhu} {and} \bibinfo{person}{Minxue
  Pan}.} \bibinfo{year}{2019}\natexlab{}.
\newblock \bibinfo{title}{Automatic Code Summarization: A Systematic Literature
  Review}.
\newblock
\newblock
\showeprint[arxiv]{1909.04352}~[cs.SE]


\end{thebibliography}
